\documentclass{myar2e}

\pdfoutput=1

\usepackage{graphics,graphicx,aasdefs,epsfig,natbib,amsmath}

\newcommand{\sigmav}{$\langle\sigma v\rangle$}

\newcommand{\egret}{{EGRET}}

\newcommand{\fermilat}{\textit{Fermi}-LAT}

\newcommand{\pamela}{{PAMELA}}
\newcommand{\pamelalong}{{Payload for Anti-Matter Exploration and Light-nuclei Astrophysics}}
\newcommand{\iact}{{IACT}}
\newcommand{\iactlong}{{Imaging Air Cherenkov Telescope}}
\newcommand{\hess}{{HESS}}

\newcommand{\magic}{{MAGIC}}

\newcommand{\veritas}{{VERITAS}}

\newcommand{\icecube}{{IceCube}}
\newcommand{\wmaplong}{{Wilkinson Microwave Anisotropy Probe}}
\newcommand{\wmap}{{WMAP}}

\newcommand{\url}{{\rm}} 

\def\Xco{$X_{\rm CO}$}
\newcommand{\hi}{H~{\sc i}}

\newcommand{\arnps}{{\it Annu.\ Rev.\ Nucl.\ Part.\ Sci.}} 
\newcommand{\jcap}{{\it J.\ Cosmol.\ Astropart.\ Phys.}}
\newcommand{\jphysg}{{\it J.\ Phys.\ G}}
\newcommand{\nima}{{\it Nucl.\ Instrum.\ Methods\ Phys.\ Res.\ A}}
\newcommand{\aph}{{\it Astropart.\ Phys.}}
\newcommand{\jhep}{{\it J.\ High Energy Phys.}}
\renewcommand{\apj}{{\it Ap.\ J.}}
\renewcommand{\apjl}{{\it Ap.\ J.\ Lett.}}
\renewcommand{\apjs}{{\it Ap.\ J.\ Supp.}}
\renewcommand{\aap}{{\it Astron.\ \& Astrophys.}}
\renewcommand{\prl}{{\it Phys.\ Rev.\ Lett.}}
\renewcommand{\prd}{{\it Phys.\ Rev.\ D}}
\renewcommand{\mnras}{{\it MNRAS}}
\renewcommand{\araa}{{\it Annu.\ Rev.\ Astron.\ Astrophys.}}

\def\simlt{\mathrel{\hbox{\rlap{\hbox{\lower4pt\hbox{$\sim$}}}\hbox{$<$}}}}
\def\simgt{\mathrel{\hbox{\rlap{\hbox{\lower4pt\hbox{$\sim$}}}\hbox{$>$}}}}

\makeatletter
\renewcommand{\section}{\@startsection{section}{1}{\z@}%
                                   {-3.5ex \@plus -1ex \@minus -.2ex}%
                                   {2.3ex \@plus.2ex}%
                                   {\normalfont\large\bfseries}}
\makeatother

\begin{document}

\jname{Annu.~Rev.~Astron.~Astrophys.}
\jyear{2011}
\jvol{49}

\title{Dark Matter Searches with Astroparticle Data}

\author{T.~A.~Porter,$^1$ R.~P.~Johnson,$^2$ and P.~W.~Graham$^3$
  \rule[-0.4cm]{0cm}{0.7cm}
  \affiliation{[1] Hansen Experimental Physics Laboratory,
    Stanford University, Stanford, California 94305; email: tporter@stanford.edu
\vspace*{0.5\baselineskip}\newline
  [2] Santa Cruz Institute for Particle Physics, University of
  California, Santa Cruz, California 95064; email: rjohnson@scipp.ucsc.edu
\vspace*{0.5\baselineskip}\newline
  [3] Stanford Institute for Theoretical Physics, Stanford
  University, Stanford, California 94305; email:
  pwgraham@stanford.edu}}


\begin{abstract}
The existence of dark matter (DM)
was first noticed by Zwicky in the 1930s, but
its nature remains one of the great unsolved problems of physics.
A variety of observations indicate that it is non-baryonic and
non-relativistic.
One of the preferred candidates for non-baryonic DM is a weakly
interacting massive particle (WIMP) that in most models is stable.
WIMP self-annihilation can produce cosmic rays, gamma rays, and
other particles with signatures that may be detectable.
Hints of anomalous cosmic-ray spectra found by recent experiments, such
as PAMELA, have
motivated interesting interpretations in terms of DM annihilation and/or decay.
However, these signatures also have standard astrophysical
interpretations, so additional evidence is needed
in order to make a case for detection of DM annihilation or decay.
Searches by the \fermilat\ for gamma-ray signals from clumps, nearby
dwarf spheroidal galaxies, and galaxy clusters have also been performed, along
with measurements of the diffuse Galactic and extragalactic gamma-ray emission.
In addition, \iactlong{s} like \hess, \magic, and \veritas\ 
have reported on searches
for gamma-ray emission from dwarf galaxies.
In this review, we examine the status of searches for particle DM
by these instruments and discuss the interpretations and resulting DM limits.
\end{abstract}

\maketitle


\section{INTRODUCTION}
\label{sec_introduction}

The combination of many observations, including galactic rotation curves,
gravitational lensing, the cosmic microwave background (CMB), and
primordial light element abundances, cannot be explained without
extending the standard model of particle physics.
The simplest extensions involve dark matter (DM) composed of a new particle
that may annihilate or decay to standard model particles 
detectable far from their source.
Although there are many other motivations for physics beyond the standard model,
astrophysics observations provide some of the few pieces of direct
evidence that there must be new physics, thus making the
search for signatures of particle DM an especially compelling area of research.

In this review, we examine recent astroparticle 
experiments and data that seek
to discover the particle nature of DM and determine its
properties indirectly through the detection of cosmic ray (CR), gamma ray, and 
neutrino signatures. 
Such work is complementary to direct and accelerator-based searches, because
it is unlikely that enough information will be obtained from a single 
method alone to determine all of the DM properties.
For example, accelerators may detect new candidate particles but 
cannot ascertain whether they form the DM.
Direct and indirect detection experiments rely on different unknown 
properties of particle DM, so it is important to pursue both.
See, e.g., \cite{Baltz:2006fm} 
and \cite{Hooper:2008sn} for comprehensive discussions 
of how the DM particle sector might eventually be understood
by following these complementary avenues of research.
DM searches with astroparticle data have the potential to 
determine the astrophysical distribution of the DM particles, which is 
not possible with the other methods.

We give a general overview of DM characteristics and popular
models, and discuss the recent astroparticle experimental searches for
evidence of its particle nature.
However, the subject has a large literature and we cannot
possibly cover it in entirety here.
We refer readers to the extensive reviews by \citet{Jungman1996},
\citet{Bergstrom2000}, \citet{Bertone2005}, and \citet{Feng2010} 
(and references therein) for additional background information.

\subsection{General Properties of Dark Matter}

The approximate distribution
of DM in our Universe can be deduced from its gravitational effects, but its
nature and microphysical properties are still unknown.
There are constraints on the properties of DM particle candidates
that strongly favor or rule out various models.
Non-gravitational interactions between DM and standard model
particles are highly constrained by the lack of observations of
particle DM. 
This strongly disfavors DM that is electrically charged or interacts 
by the strong nuclear force.
In addition, DM must clump gravitationally to form galaxies.
This requires DM to be ``cold'', that is, nonrelativistic at the time of
structure formation, or possibly ``warm''.
But ``hot'' DM, that is, relativistic during structure formation, cannot
explain the ensemble of data (although some hot DM
certainly exists in the form of neutrinos).
Within these constraints, the theoretically
best-motivated candidates for a DM particle are a
weakly interacting massive particle (WIMP) or an
axion. (See \cite{Feng2010} for a recent review of some other candidate
particles.)

The axion was originally postulated as a solution to the strong CP 
problem \citep{Peccei:1977hh,Wilczek:1977pj,Weinberg:1977ma}.
The review by \citet{Asztalos:2006kz} discusses axion DM, and we refer
readers to that.
We do not discuss axions in this review, except to note that one
of the few current experiments that can detect axion DM
is the Axion Dark Matter eXperiment (ADMX).
Unfortunately this instrument
will explore only a fraction of the possible axion
parameter space \citep{Asztalos:2009yp}.

Typically WIMPs are considered to be thermal relics left over
from the early Universe.
The interactions of WIMPs with standard model
particles kept them in thermal equilibrium at the high temperatures
that existed at that time.
As the Universe expanded the rate of these interactions, formation and
annihilation, eventually became too low, and the 
WIMP abundance froze out.
Thereafter their total number in the Universe no longer changed
significantly. 
(Even a decaying WIMP would have a lifetime bounded
to be much longer than the age of the universe, so only a negligible
fraction would have decayed by now.)
Therefore, the abundance today is inversely proportional
to the WIMP self-annihilation cross section.
The fractional abundance, relative to the critical density $3H^2/8\pi G$, is

\begin{equation}
\label{eqn:relic_abudance}
\Omega_{\rm WIMP}  \approx {0.1\over h^2} \left( \frac{3 \times 10^{-26} \, {{\rm cm}^3}\,{{\rm s}^{-1}} }{\left< \sigma v \right>} \right)
\end{equation}

\noindent
where \sigmav\ is the DM annihilation cross section times
the relative velocity of two WIMPs averaged over their velocity
distribution,
and $0.1/h^2$ is the approximate observed
abundance of DM \citep{Komatsu:2010fb}.
Note that this depends
on the DM annihilation cross section and fundamental
constants but not on a DM particle mass.

Therefore, to reproduce the
observed DM density of our Universe, a WIMP of any mass must have an
annihilation cross section of 

\begin{equation}
\label{eqn:standard_cross_section}
\left< \sigma v \right> \approx 3 \times 10^{-26} \, {\rm{cm}^3}\,{\rm{s}^{-1}}.
\end{equation}

\noindent
(There is some small
model-dependency in the precise cross section that yields the observed
DM abundance, but it is always quite close to this value.)
Because this is a typical value expected for a particle
with mass near the weak scale [$\mathcal{O}(100 {\rm \, GeV})$], or with
interactions suppressed by that scale,
there is a strong motivation
to consider DM models in which the candidate particle interacts
with a weak force and has a mass around the weak scale.
The fact that the observed abundance of DM points to new physics
at the weak scale,
completely independent of particle physics motivations for new
physics at the same scale, is the so-called WIMP miracle \citep[see the review 
by][and the discussion therein]{Feng2010}.

Thus, the WIMP provides a well-motivated DM candidate.  
Its couplings to the standard model particles are weak, often literally through 
the weak nuclear force.  
It most likely has a mass near the weak scale, 
though much lighter or heavier particles are possible.  
In fact, the only real bound on a thermal relic DM particle is that 
it should be heavier than $\sim$ keV so that it is 
cold instead of hot DM.  
Also, its mass should be $\lesssim 300$ TeV to 
avoid generating more than the observed DM abundance \citep{Griest:1989wd}, 
though both these limits are somewhat model-dependent.  
However, the best motivated mass range for the WIMP is within an 
order of magnitude around $\sim 100$ GeV, so we focus on that range in 
this review.

\subsection{Dark Matter Signals}

In order for the WIMP miracle to explain the DM abundance,
the WIMP must self-annihilate into standard model particles with a 
cross section close to that given by 
Equation~\ref{eqn:standard_cross_section}.  
Importantly, in most particle physics models the annihilation cross 
section is dominantly s-wave, that is, the leading order behavior of
the cross section is $\sigma \sim 1/v$ and
hence Equation~\ref{eqn:standard_cross_section} 
is independent of the velocity.
Therefore, the current value of Equation~\ref{eqn:standard_cross_section} 
is the same as in the early Universe (when the DM abundance was set), 
allowing us to predict 
the (approximate) annihilation rate at the present epoch.
Although the annihilation rate is small now, and no
longer affects the overall WIMP abundance, it may be large enough
to be observed.
The detectable stable end products
include photons, neutrinos, electrons, protons, deuterium, and 
their corresponding antiparticles. 
Because the annihilations produce equal amounts of matter
and antimatter, antimatter is a much better signal
due to its significantly lower astrophysical backgrounds.

There is no unique DM signal prediction, because the types and quantities of CRs
and other particles produced in the annihilation
depend sensitively on the nature of a DM particle candidate
and the strengths of its
interactions with standard model particles.
However, for a particular DM model, it is possible to make predictions
that are potentially testable.
Two of the more
popular DM frameworks that have been tested are the 
minimal supersymmetric standard model (MSSM) 
and universal extra dimensions.

The MSSM \citep{Dimopoulos:1981zb} provides a solution to the standard model 
hierarchy problem---the smallness of the weak scale
relative to the Planck scale---by adding a new fundamental symmetry
to the standard model: supersymmetry.
This introduces a new partner with mass around the weak
scale for every known particle in the standard model.
Of these, the lightest supersymmetric partner is stable, making it a
good DM candidate.
The nature of the lightest supersymmetric partner 
is determined by its standard model partner,
usually the Higgs or neutral gauge boson (corresponding to the
higgsino or gaugino, respectively, or mixtures of 
those known as ``neutralinos'').
The annihilation products depend on the lightest supersymmetric partner, 
its mass, and the
mass spectrum of the other new particles in the MSSM,
leading to significant model dependence even within the MSSM framework.

Univeral extra dimensions \citep{Appelquist:2000nn} are 
a commonly discussed extra-dimensional scenario.
For this framework, there is a tower of Kaluza-Klein 
partners 
for every standard model particle.
These are the extra-dimensional momentum
modes of the standard model particles and have masses that are multiples
of the inverse size of the compact extra dimensions, usually taken to be
around the weak scale.
In most versions of this theory, the lightest Kaluza-Klein
particle is stable, making it a good DM candidate.
Its annihilation products can then be predicted for any particular model.

There is also the possibility that DM particle candidates, although
cosmologically long-lived, may nevertheless decay, analogous to proton
decay in grand unified theories \citep{Arvanitaki:2008hq}.
Assuming a similar mechanism for WIMP decay, theory suggests a lifetime
of around $10^{27}$~s.
Interestingly, for WIMP masses $\mathcal{O}(1 {\rm \, TeV})$, the rate of
decay products per unit volume in that case would be similar to that
from DM annihilation, and therefore may also be observable.

\subsection{Detectability}
\label{sec:detectability}

The detectability of a DM signal from annihilation or decay depends
on the types of standard model particles, their energies, and where 
they are produced.
Particle physics describes the source spectrum for standard model particle
species $s = \gamma, e^{\pm}, ...$
in terms of a sum over all
possible annihilation final states $f$, each with branching fraction $B_{f,s}$:

\begin{equation}
\label{eqn:PPfactor}
\Phi_s(E)=\frac{1}{4 \pi} \frac{\langle \sigma v \rangle}{2 M^2 _\chi}
\sum\limits_f {dN_f\over dE}B_{f,s}\, ,
\end{equation}

\noindent
where $E$ is the secondary particle energy,
$M_\chi$ is the WIMP mass, and $dN_f/dE$ is the production rate
per annihilation of species $f$.
Substituting
$\langle \sigma v \rangle/2 M^2 _\chi \rightarrow \Gamma/ M_\chi$,
where $\Gamma$ is the decay rate, gives the corresponding production spectrum
for decaying DM.

DM may annihilate or decay to any of the standard model particles: 
quarks ($u$, $d$, $c$, $s$, $t$, $b$), 
leptons ($e$, $\mu$, $\tau$, $\nu_e$, $\nu_\mu$, $\nu_\tau$), 
or gauge bosons ($W$, $Z$, gluon, photon).  
For annihilations the final state is often a particle and its antiparticle, 
though for either annihilations or decays the final state can be more 
complicated, e.g., three-body.  
Most of these particles decay rapidly, leaving only the few stable 
particles and their antiparticles: photons, 
protons and antiprotons ($p$/$\bar{p}$), electrons and 
positrons ($e^\pm$), 
and the three flavors of neutrino ($\nu_e$, $\nu_\mu$, $\nu_\tau$).  
Note that it may also be possible to produce 
a deuterium or anti-deuterium nucleus ($D$/$\bar{D}$), which may be 
detectable with the Advanced Magnetic Spectrometer \citep[AMS;][]{Aguilar2002}
when finally deployed.  
Theoretically, each different possible final state can produce a 
different experimental signature. 
Even if only two-body final states are counted (particle-antiparticle) this 
would be 16 possible final states.  
However, in practice, most of the final states produce similar 
signatures and are often grouped together when setting experimental limits.  
So as an approximation 
it is often sufficient to consider only five types of final states.
First, those that contain quarks or the $W$, $Z$, or gluon.  
All these decay through quantum chromodynamical processes, ultimately 
producing hadrons: $p$/$\bar{p}$, and pions (also, possibly $D$/$\bar{D}$).  
The $\pi^0$s decay to gamma rays, while the $\pi^\pm$ decays produce $e^\pm$.
Second, are final states with $e^\pm$s or $\mu^\pm$s, which 
dominantly produce a hard $e^\pm$ spectrum, with the $\mu^\pm$ decays also 
producing $\nu_\mu$ and $\nu_e$.  
Third, are final states with $\tau^\pm$. 
These produce a softer $e^\pm$ spectrum and a strong neutrino signal.  
In addition, the $\tau^\pm$ can decay hadronically to 
pions (but, never protons) and thus can also produce a 
strong gamma-ray signal.  
Fourth, if there is a photon in the final state it produces a 
strong gamma-ray signal with a hard spectrum and often either a 
sharp edge, or in the case of a two-body final state, a line in the 
gamma-ray spectrum.  
Lastly, final states with neutrinos dominantly produce only a 
hard neutrino spectrum as they do not decay.  
This is summarized in table~\ref{Tab:finalstates}.  
Secondarily, each of these final states can also produce 
every other type of particle including gamma rays and protons by 
quantum loop corrections, final state radiation, and 
inverse-Compton (IC) scattering.  
However, this is at a much smaller level than the dominant 
production modes \citep[see, e.g.,][for a 
consideration of such processes in the context of DM decays]{Arvanitaki:2008hq}.

\begin{table}
\begin{center}
\begin{math}
\begin{array}{|c|c|}
\hline
\text{Final State} & \text{Dominant Signals} \\
\hline
W^\pm \text{, } Z \text{, gluon, } \text{quarks (} u \text{, } d \text{, } c \text{, } s \text{, } t \text{, } b \text{)} & p \text{, } \bar{p} \text{, } D \text{, } \bar{D} \text{, } e^\pm \text{, } \gamma \text{, }  \nu \\
\hline
e & e^\pm \\
\hline
\mu & e^\pm \text{, } \nu \\
\hline
\tau & e^\pm \text{, } \gamma \text{, }  \nu \\
\hline
\gamma & \gamma \\
\hline
\nu & \nu \\
\hline
\end{array}
\end{math}
\caption[Detectors]{\label{Tab:finalstates} 
The first column shows the possible standard model particles that DM could 
annihilate or decay into.  
The second column shows the dominant indirect detection 
signals that arise from these final states.  
At a smaller level almost every possible signal can be produced by 
any final state through loop corrections, final state radiation, and 
inverse-Compton scattering, as discussed in the text.}
\end{center}
\end{table}

If only a specific model (e.g., a supersymmetric or an extra-dimensional model) 
is considered, then the experimental limits on DM annihilation or decay 
can be derived for the particular final states predicted by that model, 
according to Equation~\ref{eqn:PPfactor}. 
However, when setting general experimental limits it is preferable to 
express them for a few representative final states, so that 
they are as model-independent as possible.  
For gamma-ray or CR detectors, it is common to consider limits for 
gamma-ray line, $b\bar b$, and $\mu^+\mu^-$ final states.
These provide a good approximation to the likely signals
(though not for neutrino detectors) from all the different
final states as explained above and in table~\ref{Tab:finalstates}.
The $\mu^+\mu^-$ final state is often motivated by models designed to 
explain the
PAMELA (\pamelalong) positron fraction data 
(see Section~\ref{sec:cosmicrays}) which require
a hard positron spectrum without significant antiproton
production.  
Limits on the $b\bar b$ final state apply to most supersymmetric models 
since it is one of the leading tree-level annihilation channels. 

Gamma rays that are produced in the DM annihilations or decays 
are undeflected by magnetic fields and travel 
to us from anywhere in the Galaxy, and
indeed almost anywhere in the visible Universe, 
effectively indicating the direction to their source. 
For a DM source of gamma rays in the nearby Universe, the flux from DM
annihilation is given by the integral of the
DM density-squared along
the line of sight from the observer to the source, multiplied by the production
spectrum:

\begin{equation}
\label{eqn:dwarfFlux}
\phi_\gamma(E,\psi)=  J(\psi)\times\Phi_\gamma(E)\, ,
\end{equation}

\noindent
where $E$ is the gamma-ray energy,
$\psi$ is the elongation angle with respect to the center of the
source, $\Phi_\gamma(E)$ is given by Equation~\ref{eqn:PPfactor}, and
the astrophysical factor is 

\begin{equation}
\label{eqn:astrofactor}
J(\psi)=\int_{\rm l.o.s.}\rho^2(\ell)\,d\ell
\end{equation}

\noindent 
where $\rho(\vec{r})$ is the density of DM particles, and 
the integration is in the direction $\psi$ along the line $\ell$. 
For gamma rays from decaying DM, $\rho^2 (\vec{r}) \rightarrow \rho (\vec{r})$
in Equation~\ref{eqn:astrofactor}.

Neutrinos interact so weakly that they
simply free-stream from anywhere in the visible Universe and indeed may
even come from the early Universe, although in that case they are 
highly red-shifted.
But, if a DM source of neutrinos is not significantly red-shifted, 
Equation~\ref{eqn:dwarfFlux} also applies for calculating the flux.

However, if the standard model particles resulting from DM 
annihilation or decays
are charged CRs, they do not travel directly to us.
Instead, they are transported
to the Solar System via scattering on magnetic irregularities in the
interstellar medium (ISM) and halo surrounding the Galaxy.
Their trajectories are quickly randomized by such processes
so that they retain little information about their initial directions.
For energies $\gtrsim 10$ GeV, the energy losses of the CR nuclei are
strongly suppressed compared to the lighter electrons and positrons.
Hence, the main effect on the CR nuclei is from scattering.
Particles
produced throughout the halo, at distances of tens of kiloparsecs and further,
can reach the Solar System.
Electrons and positrons, however, are severely
affected by IC scattering on the interstellar
radiation field
(ISRF) and by synchrotron radiation from spiraling in
the Galactic magnetic field.
If produced with energies $\gtrsim 100$ GeV, they will reach the Solar
System only if their origin is within a few kiloparsecs.

Also, CRs from DM annihilation or decay can produce gamma rays
during their transport through the ISM.
The CR intensities and spectra,
together with the target distributions (ISRF, gas, magnetic field)
determine the gamma-ray flux distribution. 
This involves treating the DM distribution as a source of CRs, calculating the
distributions and spectra of these CRs in addition
to those from standard astrophysical sources, and the diffuse
emissions taking into account the detailed distributions of the gas, ISRF, 
and magentic field.
This is a more complicated calculation than using 
Equation~\ref{eqn:dwarfFlux} and
has to be done using a numerical code, such as 
GALPROP \citep[][see also http://galprop.stanford.edu]{SM1998}, 
which allows these many details to be treated.

Disentangling the DM signals from astrophysical backgrounds is not
straightforward.
Spectral information may provide the most powerful discriminator.
It is possible to have a significant branching
fraction for DM annihilation or decay into
monoenergetic photons, giving a distinctive line in the gamma-ray spectrum.
But it is more likely that we must rely on the fact that
DM annihilation or decay
produces a relatively hard (that is, falling with increasing energy slower than 
the gamma-ray spectrum of a typical astrophysical source)
continuum spectrum with a bump or edge near the WIMP mass that is on top
of the astrophysical background.
Unless there is a very nearby DM source of CRs, spectral information is 
the only method for distinguishing between a DM signal and astrophysical
origin for these particles.

However, for gamma rays and neutrino DM signals spatial 
information can also be used.
We can take advantage of the expected
shape of the Galactic DM halo, or it can be used to search
for isolated DM subhalo objects, which can appear to be extended objects
in the gamma-ray sky if sufficiently large and nearby.
But annihilation depends so strongly on the DM spatial distributions
at small scales that our uncertainty in them
becomes a significant issue.
Massive many-body calculations are commonly used to predict the
DM distributions on
Galactic \citep[e.g.,][]{Diemand07} and extragalactic
scales \citep[e.g.,][]{MilleniumII}, but they suffer from two serious
limitations.
First, the effects of baryons are not fully
included, if at all, but in reality baryons must affect significantly the DM
distribution in important regions such as the Galactic center (GC).
Second, even the purely gravitational DM calculations cannot probe
to small scales, due to numerical limitations.
So analytic extrapolations are needed to predict behavior in the
densest regions, just
where the annihilation is most pronounced.
The resulting uncertainty is often expressed in terms
of a boost factor, which gives
the relative increase in overall annihilation rate due to
small-scale structure not predicted by the numerical many-body calculation.

\subsection{Other Indirect Dark Matter Signals}

It is also possible that annihilation of
DM trapped within the Sun or the Earth is a significant source of neutrinos.
The neutrinos would
escape and might be seen in a detector \citep{Silk:1985ax} such as
Super-K \citep{Desai:2004pq} or \icecube\ \citep{Abbasi:2009uz,Abbasi:2010}.
Such a
point source with a hard spectrum would be convincing evidence that
the neutrinos originated from DM.
DM builds up in the centers of the Earth and Sun when a WIMP
traveling through either body collides with a nucleus and
loses enough energy to become gravitationally bound.
It then orbits
the center of the object, undergoing multiple collisions.
The neutrino production rate depends on both the
DM annihilation cross section and the DM-nucleon
scattering cross section.
The latter is constrained by direct
detection experiments such as CDMS \citep{:2010zzc} and
XENON100 \citep{Aprile:2010um} to
be $\lesssim 3 \times 10^{-44} \, \rm{cm}^2$ for DM
masses $\sim 100$ GeV.

Recent work \citep{MW2007,Hooper:2010es,McCullough:2010ai} has also brought to
light the possibility that DM may be captured inside
white dwarf stars at high rates.
In fact, in certain circumstances the
energy deposited in the star by DM annihilations can
dominate its luminosity, providing a constraint on the DM-nucleon
scattering cross section.
The limit placed so far is fairly high and thus generally
constrains only models of
DM that somehow avoided the direct detection constraints.
It becomes
more interesting if the density of DM surrounding the white dwarf
is much larger than the local DM density.
Although such places may
exist, for example in globular clusters, it is difficult to know their
actual DM density.
Therefore, this limit is 
uncertain until the DM density can be measured accurately.

There is also evidence from the INTEGRAL/SPI instrument for an
anomalously large intensity of the 511-keV positron annihilation line
from the central
region of the Galaxy \citep{Teegarden:2004ct}.
Although this may
well have a standard astrophysical explanation, DM has also been
postulated as the source.
For example, light WIMPs \citep{Hooper:2008im} could annihilate
to produce positrons, or
WIMPs might upscatter to an excited state \citep{Finkbeiner:2007kk} and
then decay to produce the positrons.
See the review by \cite{revmodphyspositrons} for an extensive discussion
of possible origins of the 511-keV line.

\section{EXPERIMENTS}
\label{sec_experiments}

DM annihilation or decay into standard model particles
produces CRs, photons, and neutrinos.
CRs and gamma rays have been measured by many experiments, but the
detection of high-energy neutrinos from extraterrestrial sources has
so far proved elusive.

The experiments dedicated to the study of charged particle
CRs range from deep space probes such as NASA's {\it Advanced
Composition Explorer} \citep[ACE;][]{Stone98}, to orbiting particle detectors
such as \pamela\ \citep{Picozza07}, to massive balloon payloads, e.g.,
BESS \citep{Shikaze2007}, to enormous ground-based arrays such
as the Pierre Auger Observatory \citep{Abraham2004}.
Gamma-ray instruments also operate in space, such as
\egret\ \citep{Thompson93} on the {\it Compton Gamma Ray Observatory},
the {\it Fermi Large Area Telescope} \citep[\fermilat;][]{Atwood09},
and AGILE \citep{Tavani08},
and on the ground, e.g., the \iactlong{s} (\iact{s}) High Energy Stereoscopic
System \citep[\hess;][]{Hinton04},
Major Atmospheric Gamma Imaging Cherenkov
telescope \citep[\magic;][]{Ferenc05}, and
Very Energetic Radiation Imaging Telescope Array System
\citep[\veritas;][]{Weekes02}.
Meanwhile, \icecube\ \citep{Halzen10} is nearing completion and is anticipated
to inaugurate the era of neutrino astronomy.
Here, we provide a brief history of CR detectors and measurements
that have motivated the
search for particle DM and discuss in more detail the modern CR, gamma-ray,
and neutrino
experiments that are directly relevant to astroparticle DM searches.

\subsection{Cosmic-Ray Instruments: 1960s to Early 2000s}
\label{sec:cosmicrays1960to2000}

The first detection of CR antimatter involved
positrons \citep{DeShong1964}.
However, it was early measurements of CR antiprotons
that inspired models for production by exotic processes.
The first measurements of the antiproton/proton ratio
as a function of kinetic energy came from
balloon-borne experiments in the late
1970s \citep{Golden1979,Bogomolov1979}.
Early calculations for both antiprotons and positrons focused on secondary
production by inelastic collisions of CR nuclei with interstellar gas,
which is considered to be the standard astrophysical production
mode.
See, e.g., \cite{Shen1968} and \cite{Gaisser1974} regarding antiprotons
and, e.g., \cite{Protheroe1982} and references therein regarding positrons.
The kinematics of antiproton production combined with a steeply
falling incident CR proton
spectrum produce a distinctive spectral shape peaking
around $\sim 2$ GeV.
A large excess over the expected astrophysical background
at low energies was measured by a balloon-borne
detector \citep{Buffington1981}, which employed a different
detection method than the earlier experiments.
This stimulated interest in alternative explanations, including
the annihilation of DM in the Galactic
halo \citep[e.g.,][]{Silk1984,Stecker1985,Jungman1994}.
Annihilation signatures via other modes, e.g.,
positrons and gamma rays, were
also predicted \citep[e.g.,][]{Turner1986,Rudaz1988,Kamionkowski1991}.

Background contamination was a serious issue for these
early experiments
(see \cite{Tarle2001} and the discussion of \cite{Moskalenko2002} in
their Section 4).
Later experiments utilizing modern methods for particle identification
reduced the upper limits on the low-energy antiproton
flux \citep[e.g.,][]{Salamon1990}
until finally detection was achieved at a level 40 to 100 times lower 
than claimed by
\cite{Buffington1981}.
Prior to the launch of \pamela,
balloon-borne instruments employing differing detection methods,
such as IMAX \citep{Mitchell1996},
CAPRICE \citep{Boezio1997,Boezio2001},
BESS \citep{Moiseev97,Maeno2001,Asaoka2002}, and HEAT-pbar \citep{Beach2001}
have yielded
reliable data on the antiproton spectrum and fraction 
over the range of $\sim 200$ MeV to $\sim 50$ GeV.

Balloon experiments through the 1980s measured CR electrons and positrons
with inconsistent results \citep[e.g.,][and references therein]{Muller1990}.
Some authors interpreted the data $\gtrsim 10$ GeV for the CR positron
fraction as evidence for nearby high-energy
CR sources \citep[e.g.][]{Aharonian1995} that provided additional electrons
and positrons on top of the standard contribution by CR nuclei colliding with
the interstellar gas. 
(The 
production spectrum for the secondary positrons and electrons 
in the ISM follows the parent CR nuclei spectrum, which has a 
power-law index $\sim -2.7$. 
The primary CR electrons produced in, e.g., supernova remnants, 
outnumber the secondaries by a factor $\sim 10$ or more, depending 
on energy.
The primary electrons
have source spectra typically flatter than the secondary 
positrons by $\sim 0.3-0.5$ dex.
While propagation and energy losses further steepen the spectra, 
these effects are independent of charge sign, so the 
positron fraction from this process 
falls with increasing energy approximately as 
the ratio of the source spectra.
CR nuclei interacting with gas in the ISM cannot explain a positron 
fraction that rises with increasing energy.)
However,
the overwhelmingly large CR proton flux (protons outnumber electrons by 
a factor $\sim 100$ at GeV energies) presents a serious impediment
to accurate measurement of positrons.
The two look nearly identical in a magnetic spectrometer, so
additional instruments such as calorimeters and transition 
radiation detectors must be employed to distinguish them.
Because positrons comprise $\lesssim 0.1$\% of the total CR 
flux above $\sim 1$ GeV, 
to achieve a signal-to-noise ratio better than unity already requires
discrimination better than $\sim 10^{-3}$, which is challenging to achieve 
with controlled systematic errors.
Furthermore, because the proton spectrum falls less steeply than the 
positron spectrum, the required discrimination increases with energy.
The early experiments most likely did not achieve the necessary
level of background rejection.

The use of modern particle-physics instrumentation on balloons or in 
orbit steadily
advanced the measurement of CR positrons.
Starting from the mid-1990s, the HEAT experiment
\citep{Barwick1995,Barwick1997,Barwick1998,DuVernois2001,Beatty2004} and
CAPRICE \citep{Barbiellini1996,Boezio2000} measured the CR positron
spectrum and fraction up to $\sim 50$~GeV,
finding results that were in most respects consistent with 
standard CR nuclei-ISM secondary production.
However, a small
excess in the positron fraction above $\approx 7$~GeV was detected by HEAT
and also seen in CAPRICE data, as well as by the test
flight of AMS (Advanced Magnetic Spectrometer) \citep{Aguilar2002}.
Several possible origins for this excess were proposed, e.g., DM, pulsars,
and CRs interacting with giant molecular 
clouds \citep[e.g.,][]{Coutu1999,DuVernois2001}.
We will discuss this further in Section~\ref{sec:cosmicrays}.

\subsection{Current Cosmic-Ray and Gamma-Ray Experiments}
\label{sec:currentCRs}

Charged CRs are by far the most numerous of the high-energy
particles observed and
can be detected over a wide range of energies both
by balloon- and space-based instruments, as well as from the ground at
sufficiently high energy.
Cosmic gamma rays can be viewed only from space over much of the
spectrum of interest here, but above about 100 GeV the showers
produced by their interactions with the atmosphere can be viewed
from the ground via Cherenkov light produced by the relativistic
particles.
Both space-based and ground-based gamma-ray telescopes must
contend with a background of CRs that are more
numerous than the gamma rays by factors of $10^4$ or more.
In practice, ground-based telescopes have achieved the necessary 
signal-to-background
ratio only in small fields of view, thus restricting them to pointed
observations. 
(The Milagro experiment \citep{Atkins03}
detected gamma rays as well as CR shower products at ground level over a
very wide field of view but nevertheless did not achieve sensitivity to 
localized sources of gamma rays comparable to that of the \iact{s}.)
Here we briefly describe the CR and gamma-ray experiments currently at the
forefront of indirect searches for DM.

\subsubsection{Balloon-Borne Experiments}
\label{sec:balloons}

The Advanced Thin Ionization Calorimeter (ATIC) experiment reported in
2008 a measurement of the high-energy CR electron flux that
attracted a lot of attention (see Section~\ref{sec:cosmicrays}) due to a 
significant bump in the spectrum around 600~GeV \citep{Chang08b}.
ATIC flew on a high-altitude balloon in two Antarctic circumpolar flights
in 2000/2001 and 2002/2003 for a total exposure of
3.08~m$^2$ sr days. (Preliminary data from a third
successful flight in 2007/2008 have been shown in conferences to support earlier
results, e.g., an unpublished talk by J.~P.~Wefel at the HEAD 2010 meeting at
Waikoloa Village, Hawaii.)
The instrument is a calorimeter that was optimized for detection and
identification of CR nuclei.
It consists of a low-Z (1.2 interaction lengths of carbon)
active target, designed to
measure the charge magnitude of the incoming CR and to
initiate the first interaction, followed by a
thick (18 radiation lengths), finely
segmented bismuth germanate (BGO) calorimeter \citep{Guzik04}.
While the active target is not ideal for electron identification, the
BGO calorimeter does excel at measuring electromagnetic showers, and
the ATIC electron-proton separation was extensively studied by
simulations and in beam-test data \citep{Chang08a}.
A smaller Antarctic balloon-borne experiment, PPB-BETS, reported
an excess in the CR data [Torii et~al., unpublished data (arXiv:0809.0760)]
similar to that seen by ATIC, although with less
statistical significance due to four times smaller exposure.

\subsubsection{Space-Based Experiments}
\label{sec:spacebased}

\pamela\ is an orbiting instrument dedicated to CR
measurements \citep{Picozza07} that was launched in 2006.
It consists of a magnetic
spectrometer, an anticoincidence system (to veto particles entering 
through the sides of the spectrometer), a time-of-flight system, an 
electromagnetic sampling calorimeter of 16.3 radiation lengths thickness, 
a shower-tail-catcher scintillator, and a neutron detector.
Its spectrometer can measure the momentum and charge sign of charged 
CR particles, with a maximum detectable rigidity (momentum per unit charge) 
of 800~GeV/$c$, while the time-of-flight system, calorimeter, and 
neutron detector serve to identify the particle type.
Its acceptance for electrons is small compared to those of ATIC and
the \fermilat\ (see below), but it is uniquely able to
separate electrons from
positrons cleanly up to 270~GeV~$c^{-1}$ momentum by means of the 
magnetic spectrometer.

NASA's \fermilat\ \citep{Atwood09}, launched in 2008, is
the preeminent gamma-ray telescope in the energy range above $\sim 100$~MeV.
It is a pair-conversion telescope, like its immediate
predecessor \egret\ \citep{Thompson93} and its
much smaller contemporary AGILE \citep{Tavani08}, launched in 2007.
Thirty-six layers of silicon-strip detectors interleaved with
tungsten foils \citep{Atwood07} pair-convert the gamma rays and track the
resulting electrons and positrons.
The tracking section is followed by a segmented CsI crystal
calorimeter that measures
the energy of the electromagnetic shower.
A veto-counter system \citep{Moiseev07}, based on segmented scintillator
tiles, helps to tag charged CRs and, together with the detailed
event reconstruction in the tracker and calorimeter, reduce that background
by at least five orders of magnitude.
The tracking resolution is generally limited by multiple scattering of the
electrons and positrons in the tungsten foils and other material, giving
a point-spread function (PSF) with 68\% containment
angles for individual photons ranging from a few degrees at 100~MeV down
to about $0.1^\circ$ for energies $\gtrsim 10$~GeV.

The \fermilat\ has an extraordinarily large field of view of 2.4 sr, thus
seeing nearly 20\% of the entire sky at any instant.
It normally operates pointing outward from the Earth, scanning the sky, and
achieves a fairly uniform exposure over $4\pi$ sr by rocking back and
forth by $50^\circ$ toward one orbital pole or the other on successive orbits.
Its all-sky view together with its excellent signal-to-noise and large
energy range, from $\sim 100$ MeV to beyond 300 GeV, make it very well
suited to searches for DM annihilation in
all types of possible sources, point-like or diffuse, Galactic or
extragalactic.

The \fermilat\ is also a very capable CR electron
detector \citep{Ormes2007}.
It has no atmospheric overburden and already has an exposure at least 200
times larger (depending on energy) than that of ATIC.
See Section~\ref{sec:cosmicrays} for a comparison of the electron 
results from the two experiments.

\subsubsection{Ground-Based Experiments}

Gamma-ray astronomy with \iact{s} was pioneered by the
{\it Whipple} telescope \citep{Cawley90}, which first coupled a pixelated
photomultiplier-tube camera to a large (10-m) optical reflector,
giving it the ability to
image the atmospheric showers and thereby reject much of the CR background.
{\it Whipple} was the first \iact\ to detect an extragalactic
source, Markarian 421, in the TeV energy range \citep{Punch92}.
The four \iact{s} that dominate the field today are based on the same technique
but extend it to arrays of larger reflectors that achieve larger effective
areas, lower thresholds, and lower background.
The \hess\ experiment in Namibia \citep{Hinton04} and
the \veritas\ experiment
in Arizona \citep{Weekes02} are similar but complement each other by
being in opposite hemispheres.
Each is an array of four 13-m-diameter 
\iact{s}. (An enormous 600~m$^2$ reflector will be added to \hess\ for its 
second phase.)
Another southern-hemisphere instrument is the CANGAROO-III array of four
10-m-diameter \iact{s} \citep{Kabuki03}, while a second
northern-hemisphere instrument is \magic, a pair of 17-m-diameter
\iact{s} located on the Canary Island of La Palma \citep{Ferenc05}.

\iact{s} have small fields of view ranging from $3.5^\circ$
(\magic, \veritas) to $5^\circ$ (\hess) and therefore most often operate by
pointing at known objects.
Scans over the sky are time consuming but have been accomplished over
limited regions, such as the \hess\ scan of the
inner Galaxy \citep{Aharonian06}.
Their observing time is limited to clear, dark nights with little or no
moonlight (less than half full), for a total typically of around 900
hours per year, and their power to reject charged CR background is roughly 100
worse than that of Fermi.
However, their effective areas are very large (up to $10^5$~m$^2$
compared with less than $\approx 0.8\,{\rm m}^2$ for \fermilat).
Their trigger energy thresholds range from about 25~GeV (\magic)
to 100~GeV (\veritas), although the threshold for spectral
reconstruction is higher (150 GeV for \veritas), providing some overlap
with \fermilat\ for the brightest sources (such as the Crab pulsar wind nebula).
In general, \iact{} energy resolutions are about 15\% at high energy, with
single-photon angular resolutions of
around 0.1$^\circ$ (similar to the resolution achieved by \fermilat\
above 10~GeV).
More details on the \iact{} method and instruments are available in the
review by \cite{Hinton09}.


\subsection{Neutrinos}

So far, neutrinos from sources outside of the Solar System have not
been detected except in the case of a single fortuitous supernova event
in the Large Magellanic Cloud,
SN1987A, from which neutrinos were detected by two underground
experiments \citep{Bionta87,Hirata87}.
However, new experiments with very large
sensitive volumes are planned or are under construction with the intent
to realize the potential of neutrino astronomy \citep{Anchordoqui2010}.

Only the $>90\%$ completed \icecube\ detector \citep{Halzen10} at the
South Pole is currently reaching the cubic-kilometer scale
thought to be necessary for the dawn of neutrino astronomy.
Neutrinos traveling through the Earth can interact to produce particles
traveling upward through the Antarctic ice.
Cerenkov light from the relativistic particles is detected by
photomultiplier tubes arranged
on 80 strings 2,450~m long descending into the ice cap, with the active
volume all below a depth of 1,450~m.

\icecube\ has an irreducible foreground of neutrinos produced in
Northern-Hemisphere CR showers.
Those are almost all muon neutrinos, so it is advantageous to
distinguish electron and tau neutrino events in \icecube\ from those
produced by muon neutrinos, even though the latter give the best
directional information.
Very high-energy tau leptons can travel hundreds of meters in the detector
before decaying, yielding distinctive double-vertex events, whereas electrons
and low-energy taus deposit their energy in localized showers,
giving excellent energy resolution but little directional information.

The angular resolution for long muon events is about one degree, and the
energy threshold is $\approx$ 100~GeV.
Six additional closely spaced strings of detectors lowered into the
deepest 350~m of ice will form a ``Deep Core'' infill array with a
significantly lower (10~GeV) threshold.
The \icecube\ detector has been operational even during construction,
and neutrino sky maps have been made with a half-year exposure of a 1/2
cubic km detector, yielding nearly 7,000 neutrinos from the Northern
Hemisphere \citep{Halzen10}.

\section{RECENT DATA AND INTERPRETATIONS}
\label{sec_recent_data}

\subsection{Cosmic Rays}
\label{sec:cosmicrays}

\begin{figure}
\begin{center}
\includegraphics[height=2.35in]{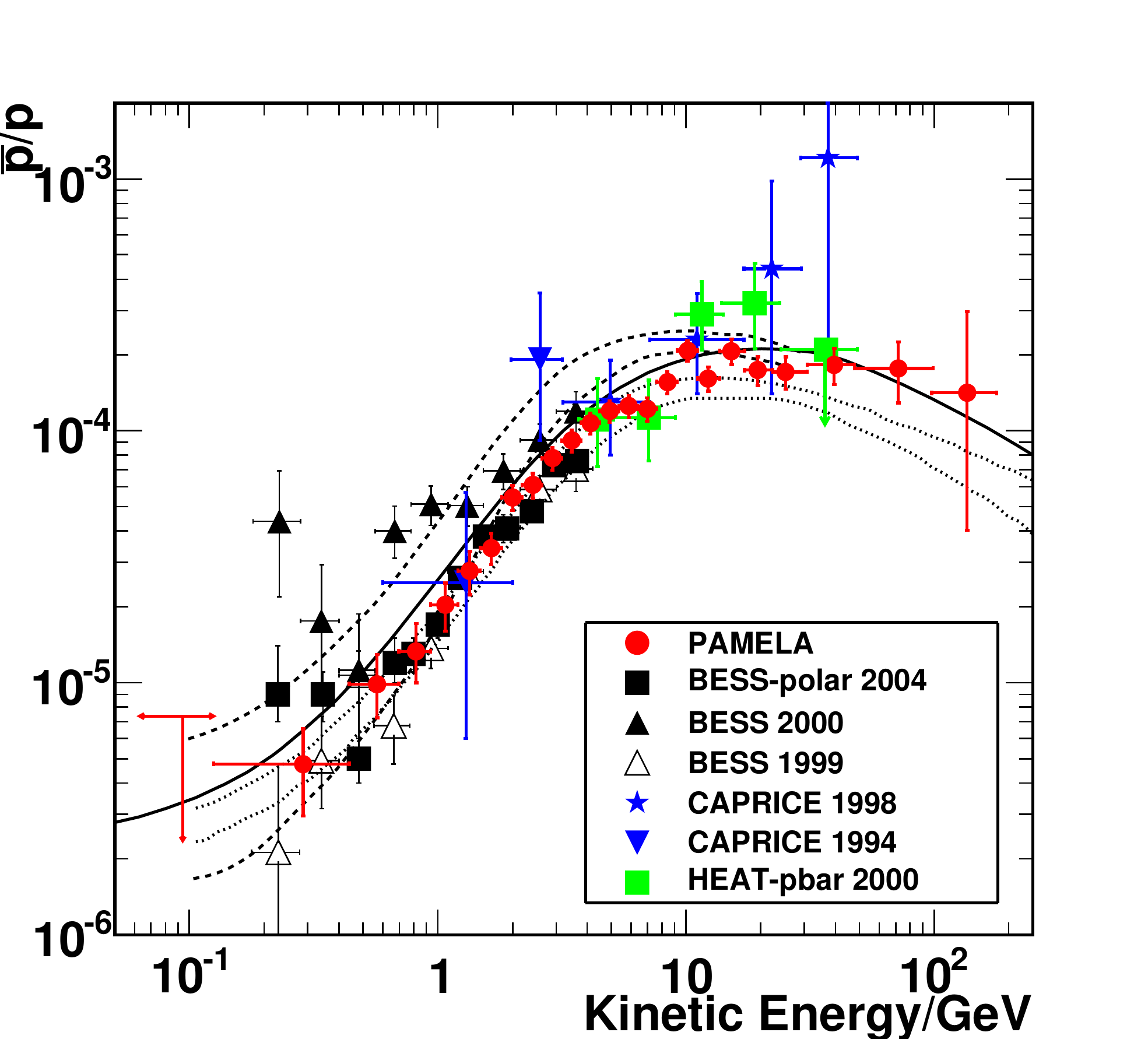}
\includegraphics[height=2.35in]{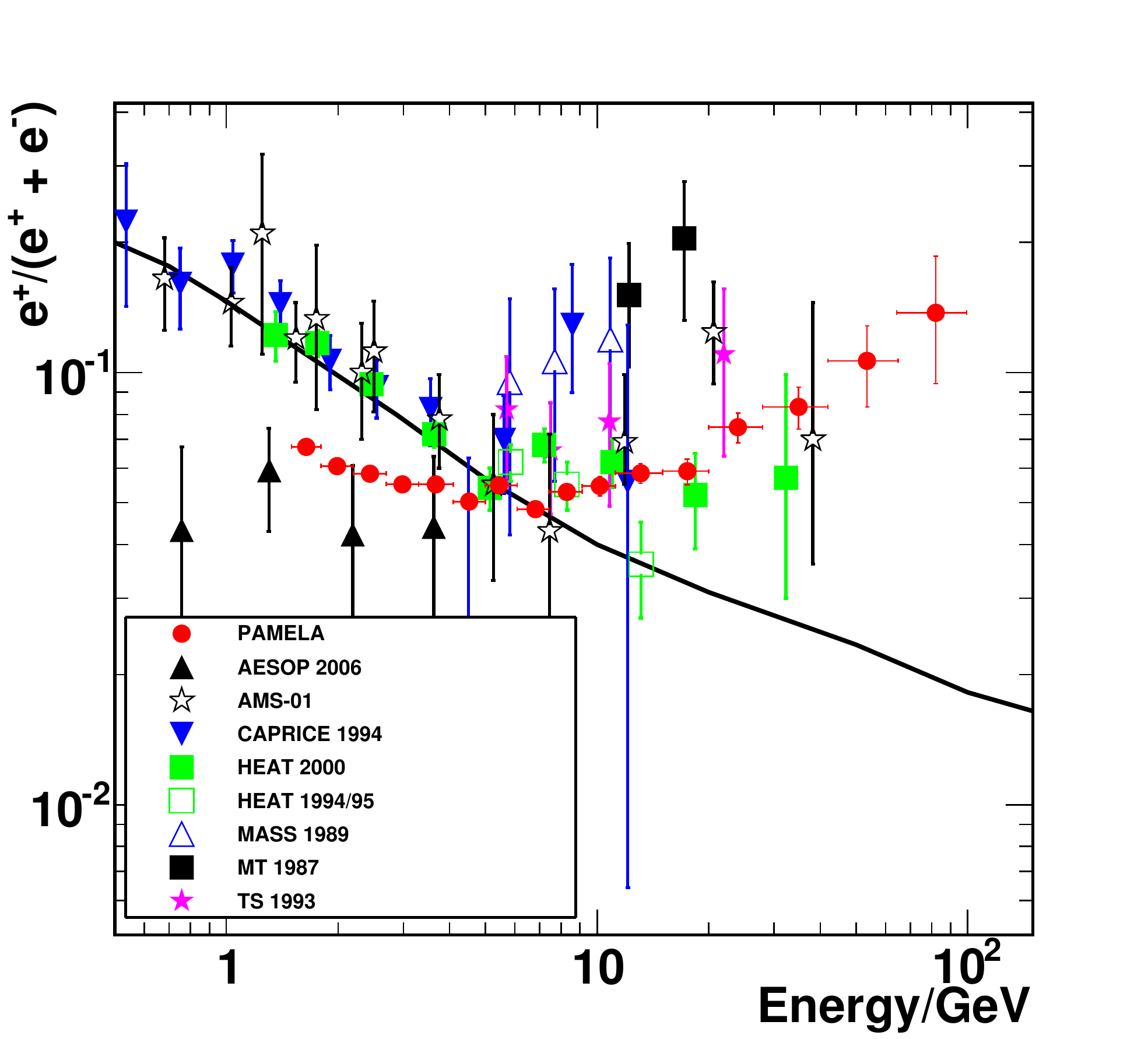}
\end{center}
\caption{\small
(a) Cosmic-ray (CR) antiproton fraction and (b) positron fraction.
The CR measurements by various instruments are summarized in 
\citet{Adriani2010pbar} (antiprotons) and \citet{Adriani09} (positrons).
For antiprotons, the curves correspond to models with different
assumptions for the treatment of CR propagation, uncertainties in the assumed
propagation model parameters, and cross section uncertainties for antiproton
production, annihilation, and scattering. 
Upper and lower dashed lines were calculated for a homogeneous 
(leaky box) model by \citet{Simon1998}.
Upper and lower dotted lines were calculated assuming a diffusive 
reacceleration with convection model by \cite{Donato2009}.
Solid line shows the calculation by \citet{Ptuskin2006} for a plain diffusion
model.
For positrons, the solid curve shows the prediction by \cite{MS1998} 
using the GALPROP code for CR nuclei interacting with the
interstellar gas for a plain diffusion model without
accounting for solar modulation effects.
Figures are adapted from original forms published in \citet{Adriani2010pbar}
and \citet{Adriani09}. 
}
\label{figCRPAMELA}
\end{figure}

The \pamela\ instrument team has reported
measurements of the antiproton spectrum and
fraction \citep{Adriani:2008zq,Adriani2010pbar}, and the positron
fraction \citep{Adriani09,Adriani2010pos}.
Figure~\ref{figCRPAMELA}a
shows the combined measurements of the CR antiproton fraction
as of late 2010 \citep{Adriani2010pbar},
whereas Figure~\ref{figCRPAMELA}b shows the
CR positron fraction as measured by many experiments
up to late 2009 \citep{Adriani09}.
The \pamela\ antiproton measurements agree with
earlier data (where there is overlap), which
are consistent with expected non-exotic astrophysical origins.
However, the \pamela\ positron fraction rises
with increasing energy, opposite to
the expected behavior of secondaries produced in the ISM (see
Section~\ref{sec:cosmicrays1960to2000}).
The \pamela\ data apparently confirm the results
from the earlier HEAT balloon experiment
and AMS test-flight (although the results of both of those experiments 
have much larger uncertainties).

An essential question for these data is the likelihood that they are
the result of an experimental artifact. (Recall,
in Section~\ref{sec:cosmicrays1960to2000} we discussed how reliable
proton-positron discrimination is essential for this measurement.)
\pamela\ uses its magnetic spectrometer, time-of-flight 
system (at low energy), calorimeter,
and neutron detector for the separation of protons and antiprotons
from positrons and electrons (see Section~\ref{sec:spacebased}).
The spectrometer separates the electrons and antiprotons from the
positrons and protons \citep[except at the highest energies, where there is some spill-over;][]{Adriani2010pos}.
The calorimeter is able to separate
electromagnetic- and hadron-initiated (proton/antiproton)
showers very well using information
on the longitudinal and lateral shower development.
However, early neutral pion production at the top of the calorimeter by
interacting hadrons produces an electromagnetic shower in hadron-initiated
events at about the percent level.
This makes the separation between true electromagnetic- and hadron-initiated
events difficult when this occurs because the two look essentially identical
in the calorimeter.
For the protons and positrons, because the ratio of these particles
is large (in favor of the protons) and rises
with energy, a slightly larger than expected misidentification of
protons could easily lead to a rising positron fraction.
The \pamela\ collaboration has published details of their analysis,
including performance of the particle discrimination \citep{Adriani2010pos}.
However, some researchers have questioned if sufficient rejection power
is obtained by the instrument \citep{Schubnell2009} because
additional hadron/electromagnetic discrimination systems, like transition
radiation detectors, are not used.
It would be useful to have the absolute positron spectrum that
\pamela\ measures to ascertain whether anomalous spectral features also exist
in that, but so far it has not been published.
These questions will be resolved by follow-up measurements by the
AMS experiment, which is due for deployment on the International Space
Station in early 2011.
AMS will employ multiple systems, such as calorimeters and
transition radiation detectors, to separate protons
and positrons \citep{Carosi2004}.
It will also separate electrons from positrons to higher energy and will have
a much larger acceptance than \pamela.

\begin{figure}
\begin{center}
\includegraphics[width=5.5in]{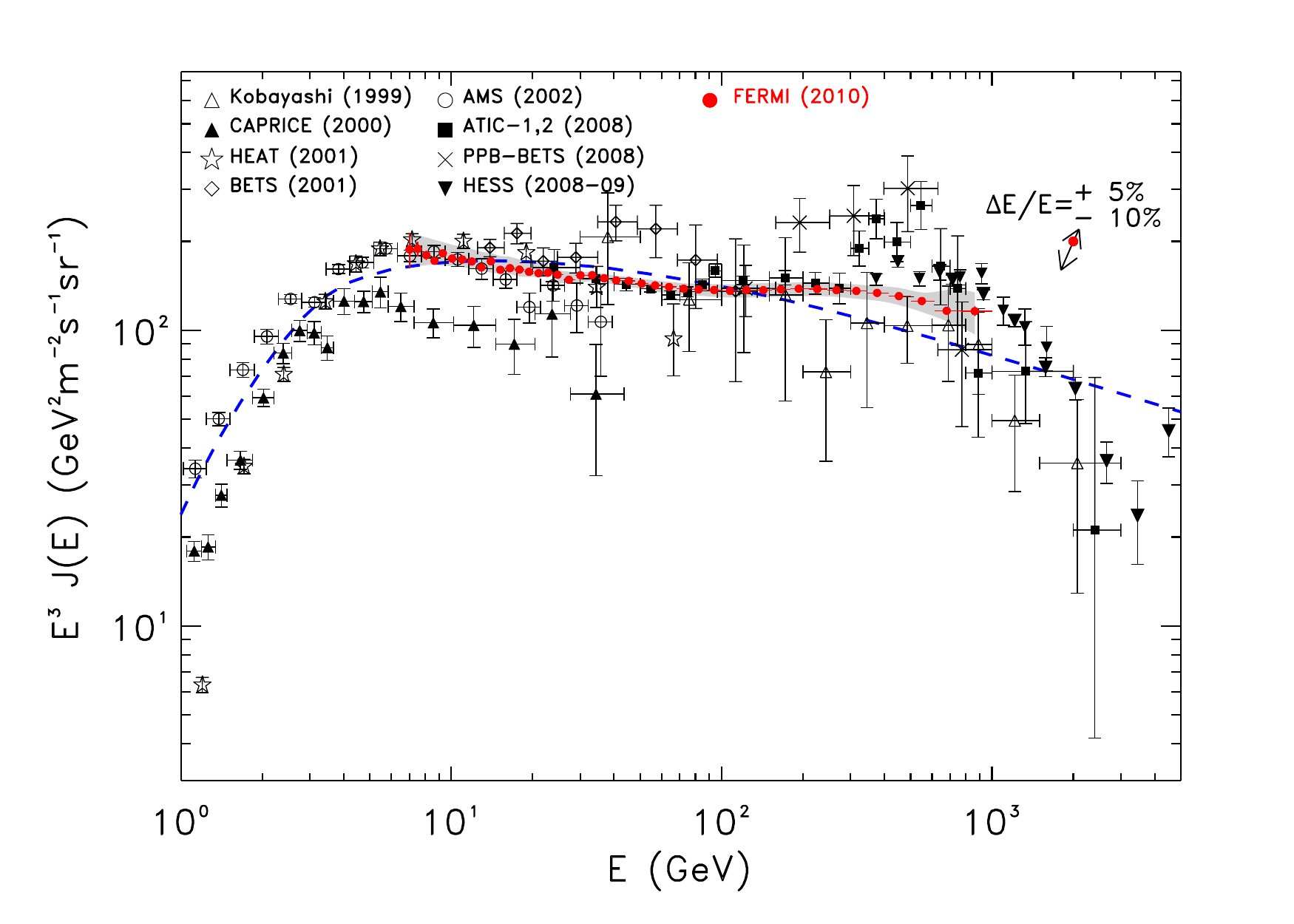}
\end{center}
\caption{\small
Combined cosmic-ray (CR)
electron and positron spectrum as measured by \fermilat\
from $\sim 7$ GeV to 1 TeV for one year of
observations, together with other measurements
\citep[][and references therein]{Abdo:2009elec1,Ackermann2010elec}.
The systematic errors for the \fermilat\ measurement are shown by the grey
band.
The systematic uncertainty associated with the absolute energy scale
is shown by the non-vertical arrow in the upper right corner of the figure.
(The arrow shows the rigid shift of the entire spectrum due to the
uncertainty in the energy scale that takes into
account multiplying the spectrum by $E^3$.)
The dashed line shows the predicted total CR electron and positron
spectrum for a diffusive-reacceleration CR propagation model based on
CRs and other data measured prior to the \fermilat\ results
\citep{Abdo:2009elec1}.
This model was not tuned to the \fermilat\ measurements and used
a CR electron injection spectrum that was a
single power law above $\sim 100$ GeV.
The figure clearly illustrates that
the data are not well-reproduced by such a simple spectral model.
Possible explanations are discussed in the text,
the most likely being that the underlying assumptions commonly
used to treat astrophysical CR sources and propagation are too
simplistic.
}
\label{figCRelectron}
\end{figure}

Recent measurements of the total CR electron spectrum have also shown
anomalies.
(We use `electron' here to mean both electrons and positrons,
since the ATIC and \fermilat\ instruments are not able to discriminate
particles on the basis of charge.
When we need to distinguish between electrons and their anti-particle, 
we explicitly refer to the latter as positrons.)
Toward the end of 2008 the ATIC collaboration \citep{Chang08b} announced
a surprising result: a bump
in the CR electron spectrum in the energy range $\sim 300-800$ GeV,
where conventional
astrophysical sources are expected to produce a smooth, power-law spectrum.
The PPS-BETS experiment
[Torii et~al., unpublished data (arXiv:0809.0760)] detected a
similar excess over approximately the same energy range as ATIC,
although with less statistical significance.
See Figure~\ref{figCRelectron} for the combined measurements of
the CR electron spectrum up to the end of 2010.
Together with the positron fraction measured by \pamela, these
results stimulated a lot of speculation about the
origin of the spectral features: DM, certainly, but also that the
assumptions for the distribution of astrophysical CR sources and propagation
were too simplistic.
[This has been suggested since the late 1960s but the measurements
until those recently by the \fermilat\ 
have suffered from large uncertainties at high energies, where the usual 
assumptions break down.
Even with the significantly improved measurements
it is still difficult to make strong conclusions about the
origin(s) of the
CR electrons measured above a few hundred GeV. We discuss this further below.]

However, as Figure~\ref{figCRelectron} shows,
the measurement by the \fermilat\ of the CR electron
spectrum from 20~GeV to 1~TeV
\citep{Abdo:2009elec1} smoothly connects with the \hess\ electron
measurements 
at higher energies \citep{Aharonian:2008elec,Aharonian:2009elec}.
The \fermilat\ measurement does not show the anomalous
spectral features described by ATIC and PPB-BETS.
Extensive details of the \fermilat\ analysis were provided in a
follow-up publication \citep{Ackermann2010elec}, which also extended the
previous measurements down to the geomagnetic cut-off
energy for the \fermilat\ orbit ($\sim 7$ GeV).
Detailed simulations of the \fermilat\ instrument and comparison with
beam-test data demonstrated that the
energy resolution was more than adequate to see a spectral feature like
the ATIC peak.
Because of the large statistics for the \fermilat\ CR electron data (over
8 million electron candidate events for the first 12 months of the mission) it
was possible to apply more stringent cuts on subsets of the data to cross
check the full analysis.
An analysis was done for a subset of events using only tracks passing
through at least 12 radiation lengths of calorimeter (16 radiation lengths
on average), which found consistent results with the analysis of the full
event sample.
This conclusively ruled out the possibility that any ATIC-like feature was 
missed in the earlier \citet{Abdo:2009elec1} analysis.

Nevertheless, there is a less dramatic feature apparent above $\sim 200$ GeV 
in the
\fermilat\ spectrum.
Its significance is hard to estimate, due to the
existence of both correlated and uncorrelated (from
energy bin to energy bin) systematic uncertainties in the instrument
acceptance.
Separating the two types of uncertainties is a difficult problem 
that so far has not been accomplished by the \fermilat\ team.
While the data are compatible with a
power-law spectrum within the displayed band of systematic uncertainties, 
if a model with a
power-law spectrum that is constrained by
other data (such as the model curve shown in Figure~\ref{figCRelectron})
is compared with the \fermilat\ spectrum, then
the significance of the spectral feature above $\sim 200$~GeV can be very high.
Therefore, the \fermilat\ data
and the rising positron fraction measured by \pamela\ have
motivated the construction of DM models to reproduce the apparent
features observed by these instruments.

Because DM
annihilation or decay creates equal amounts of matter and
antimatter with a hard spectrum up to the DM particle
mass [$\mathcal{O}(100 \, \rm GeV)$],
it can naturally produce a rising positron fraction and
has been a widely conjectured explanation of the \pamela\ positron data.
However, models must also explain several unexpected characteristics of
these data.
The relatively large number of additional electrons and positrons
produced requires a self-annihilation cross 
section $\sim 10^{2}$ to $10^{3}$ times
larger than given by Equation~\ref{eqn:standard_cross_section} (see 
Figure~\ref{figFermiDwarfLepton2}, also).
In addition, the measured antiproton fraction does not rise with
energy (see Figure~\ref{figCRPAMELA}a), which requires hadron production
to be suppressed.
Because such a large enhancement of the cross section cannot be simply
ascribed to uncertainty in the spatial distribution of the DM (the boost
factor; see Section~\ref{sec:detectability})
several kinds of non-standard DM
models have been developed to fit both the \pamela\ and \fermilat\ data.
(The \fermilat\ data provide an upper bound to enhancements in the
cross section.)
These
include adding a Sommerfeld enhancement to the cross section due to a
hypothetical long-range force \citep[e.g.,][]{ArkaniHamed:2008qn}, a nonthermal
production mechanism \citep[e.g.,][]{Nelson:2008hj,Fairbairn:2008fb},
or DM decay \citep[e.g.,][]{Arvanitaki:2008hq}.

A more likely explanation of these CR data is that the assumptions made
estimating the astrophysical background are overly simplified.
Any DM signal in CRs
exists on a background of particles from astrophysical sources
(e.g., supernovae, supernova remnants, pulsars,
compact objects in close binary systems, stellar winds, etc.).
The primary CRs, mostly nuclei, are produced at their sources,
eventually escaping to propagate
in the ISM of the Galaxy, where their spectra change over
tens of millions of years as they
lose or gain energy through interactions with the interstellar gas,
ISRF, magnetic fields and turbulence.
Their composition also changes as
the destruction of primary CR nuclei via spallation on the interstellar gas
gives rise to secondary particles,
including nuclear
isotopes that are rare in nature, antiprotons, and electrons, positrons,
and neutrinos from the decays of charged pions.
Within the Solar System, charged CRs
with energies below a few tens of GeV energies are affected by the solar wind.
This solar modulation alters the interstellar spectra to produce what
we observe locally.
Our understanding of CR sources and propagation, the distributions
of the target distributions in the ISM (interstellar gas, ISRF,
magnetic field), and the heliospheric transport of
CRs all impact our ability to disentangle DM from astrophysical signals.

For modeling CR production from astrophysical sources, the
assumption is usually made that the sources are described by a
smoothly varying function of position, and that they have a
common characteristic source spectrum.
The assumed spatial distribution is typically based on supernova remnant
and pulsar surveys
\citep[e.g.,][]{Case1998,Lorimer2006}.
That is an acceptable approximation for CR nuclei, barring
errors in the data-derived distributions, which are influenced by
selection effects.
However, for CR electrons and positrons the severe radiative
losses from interactions with the ISRF and interstellar magnetic field 
mean that a source of
high-energy electrons must be within $\sim 1$~kpc if the detected particles
are to have energies larger than a few hundred GeV
\citep[e.g.,][see also Section~\ref{sec:detectability}]{Berkey1969,Shen1970,
Atoyan95}.
Therefore, assuming a smooth spatial distribution is not correct, and
details of the discrete source distribution in the vicinity of 
the Solar System are important.
Many authors also assume that the ISM is
homogeneous, which is incorrect for these particles.
[This is typically done for analytic
treatments of the CR propagation, since it is not possible to treat
arbitrary spatial distributions for the ISM components that determine
the energy losses. Even at the highest Galactic CR energies
($\gtrsim 100$ TeV), where the dominant electron/positron energy losses are
IC scattering of the CMB and synchrotron radiation, inhomogeneities in the 
magnetic field are important.]
Furthermore, the assumption of a single source spectrum is unrealistic, because
there are a variety of production mechanisms for high-energy
electrons and positrons: acceleration at
shocks \citep[e.g.,][]{Lagage1983,Hillas2005}
(electrons only),
production of secondary electrons/positrons {\it in situ} by shock
accelerated hadrons \citep[e.g.,][]{Blasi2009}, pair production
via different mechanisms \citep[e.g.,][]{Aharonian1991,Chi1996}, in addition
to possible production by DM annihilation or decay.

Although there is no theory of CR propagation based on first principles, the
phenomenological description provided by isotropic diffusion
models (diffusion-convection, diffusive-reacceleration, etc.) has proven
very successful in describing a wide range of CR data: stable
secondary nuclei, radioactive nuclei, electrons, and so
forth \citep[e.g.,][and references therein]{Strong2007}.
In these models, the propagation parameters and boundary conditions are
obtained from fits to the CR nuclei data for the
secondary-to-primary ratios (e.g., $_5$B/$_6$C and
[$_{21}$Sc+$_{22}$Ti+$_{23}$V]/$_{26}$Fe), and radioactive
abundances (e.g., $^{10}_{4}$Be, $^{26}_{13}$Al, $^{36}_{17}$Cl,
and $^{54}_{25}$Mn).
Correcting for solar modulation is typically done using the force-field
approximation \citep{Gleeson1968} that uses a single parameter, the
modulation potential, to characterize the strength of the effect.
Direct comparison between propagation model calculations and data is
problematic because the modulation potentials from different
experiments cannot be interpreted independently from the experiments.
The derived values depend on the choices of interstellar
spectra used for their analyses, which differ from experiment to
experiment (and are sometimes not provided).
Coupled with the uncertainties in the propagation model,
cross sections, and other details, finding a unique set of parameters 
to describe the CR nuclei data is challenging, which also
affects the predictions for CR leptons.
Some recent work illustrates their effect for predictions of 
CR positron spectra using
analytic propagation model solutions \citep[e.g.,][]{Delahaye2009}.
So far, a systematic exploration using a numerical code such as GALPROP
has not been performed.

A more subtle issue is related to the treatment of the inhomogeneous
CR electron source distribution within a particular propagation model.
To our knowledge, a self-consistent calculation deriving the propagation
model parameters from the CR nuclei data, and including the discrete
electron/positron sources while taking into account the inhomogeneous energy
losses, has not been accomplished.
Researchers commonly use the free-space 3-dimensional spherically symmetric
analytic diffusion results of \citet{Aharonian1995}
to treat the nearby source contribution and then a diffusion model variant
(either numerical or analytically solved) with a smooth
spatial distribution for the far ($\gtrsim 1$ kpc)
sources (it is reasonable to assume a smooth spatial distribution for
these because the propagation and energy losses erase any details
related to the discrete distribution).
The problem is that the propagation parameters used to calculate
the nearby contribution are obtained from the propagation
model used for the far sources [which may not have the same
spatial configuration (2/3-dimensional cylindrical/cartesian),
boundary conditions, and/or distributions for ISM components], or
simply adopted from the literature.
This inconsistent treatment of the sources and propagation by essentially
all calculations done so far clouds the interpretation of their results.

Given the uncertainties associated with modeling the CR sources, propagation,
and so forth,
we believe that it is highly plausible that both the \pamela\ data and
the total electron spectrum
can be explained by modifying the assumptions usually made
when calculating the standard astrophysical background.
For example, the electron and positron data have been explained
in terms of contributions from the nearest known
pulsars and supernova
remnants \citep[e.g.,][and many other papers]{Cowsik1979,Kobayashi2004}.
But we emphasize that even for these calculations there is no
unique set of sources that have been identified to explain the
measured data.
In these models, the injected CR power by each source is unknown and is treated
as a free parameter that is adjusted to reproduce the observations. 
Depending on the assumed production mode for the CRs there are
some restrictions on inhomogeneous models,
and care must be taken not to violate other observational constraints.
For example, {\it in situ} production of secondary electrons and positrons
also produces more antiprotons and secondary nuclei if CR nuclei
are accelerated together with protons \citep{Lagage1985}, which can be
inconsistent with other measurements (but, again, such models
have additional adjustable parameters, e.g.,
the amount of matter in the CR confinement region at the source,
which directly affects the total number of
secondaries produced and injected into the ISM).
Or for pair production on soft photon fields \citep[e.g.,][]{Stawarz2010},
intensities for the ISRF much higher than observed are needed in
order to produce the required numbers of additional positrons.
In short, despite being apparently viable explanations of the 
CR data,
none of the non-exotic models can be pointed to as the unique solution.
Thus the DM explanations, although not essential, also remain viable.

\subsection{Galactic Diffuse Emission}
\label{sec:diffuseemission}

The diffuse non-thermal emission of photons from radio to gamma-rays is
closely connected to the production and propagation of CRs.
In inelastic collisions with the interstellar gas, CR nuclei
produce neutral pions,
which decay to gamma rays.
Primary CR electrons and secondary electrons and positrons
produce gamma rays via bremsstrahlung with the interstellar gas and by
IC scattering off of the ISRF.
They also
produce diffuse emission in the radio to microwave band by
synchrotron radiation induced by the Galactic magnetic field.
Because gamma rays are not deflected by magnetic fields, and because
their absorption
in the ISM is negligible over Galactic distances up to
energies of $\sim 10^5$ GeV \citep{Moskalenko2006},
they directly probe CR spectra and intensities
in distant locations (see \citet{MSR2004} and \citet{Strong2007} for reviews).
Here we discuss only the diffuse emission of the Milky Way, for which we
have the best data and which are most relevant to
constraining the contributions from DM.

The EGRET ``GeV excess'' was an anomalous signal in the diffuse Galactic
emission observed in EGRET data.
The term referred to emission for gamma-ray energies $\gtrsim 1$ GeV that
was in excess of diffuse gamma-ray emission models based on locally
measured CR spectra \citep[see, e.g., Figure~7 of][]{Hunter1997}.
It was proposed that the GeV excess was due to gamma rays
from annihilating DM \citep{deBoer2005a,deBoer2005b}.
This received much attention, but a number of more conventional
or prosaic explanations were also considered, e.g.,
variations in the CR spectra
\citep{Porter1997,SMR2000,SMR2004} and hypothesized instrumental
effects \citep{Hunter1997}.
The DM interpretation was itself challenged because the DM models employed
over-produced antiprotons relative to the data already 
available \citep{Bergstrom2006}.

Testing the origin of the GeV excess was one of the
early studies of the diffuse gamma-ray emission by the \fermilat\ team
\citep{Abdo:2009gevexcess}.
The data at intermediate
Galactic latitudes ($10^\circ \leq |b| \leq 20^\circ$) were used in the study
because the
standard astrophysical production of the diffuse gamma-ray
emission over this region of the sky come predominantly
from relatively nearby CR nuclei interactions with interstellar gas.
The majority of the gas is in the form of atomic hydrogen (\hi) and molecular
hydrogen (H$_2$), along with a small amount of ionized hydrogen. 
The \hi\ has roughly uniform density at $\sim 1$ atom cm$^{-3}$, with a
typical scale height about the Galactic
mid-plane of $\sim 200$~pc \citep{Dickey1990,Kalberla2009}.
The H$_2$ is concentrated mainly
in clouds of density typically $\sim 10^{3-4}$ molecules cm$^{-3}$ and mass
$10^{4-6}$ M$_\odot$, with a scale height of $\sim 70$~pc \citep{Combes1991}.
For these latitudes, the CRs producing the majority of the
diffuse emission are therefore several hundreds of parsecs
to $\sim 1$~kpc from the Sun.
Because the CRs are mainly nuclei, for which energy losses are slow,
the CR intensities and spectra are expected to be close to those
measured locally.
Significant spectral anomalies like the GeV excess would then stand out
against the standard astrophysical signal, because
this region of the sky avoids
the more model-dependent complications inherent in
understanding the emission closer to the Galactic plane or at higher
latitudes.

\begin{figure}
\begin{center}
\includegraphics[width=2.6in]{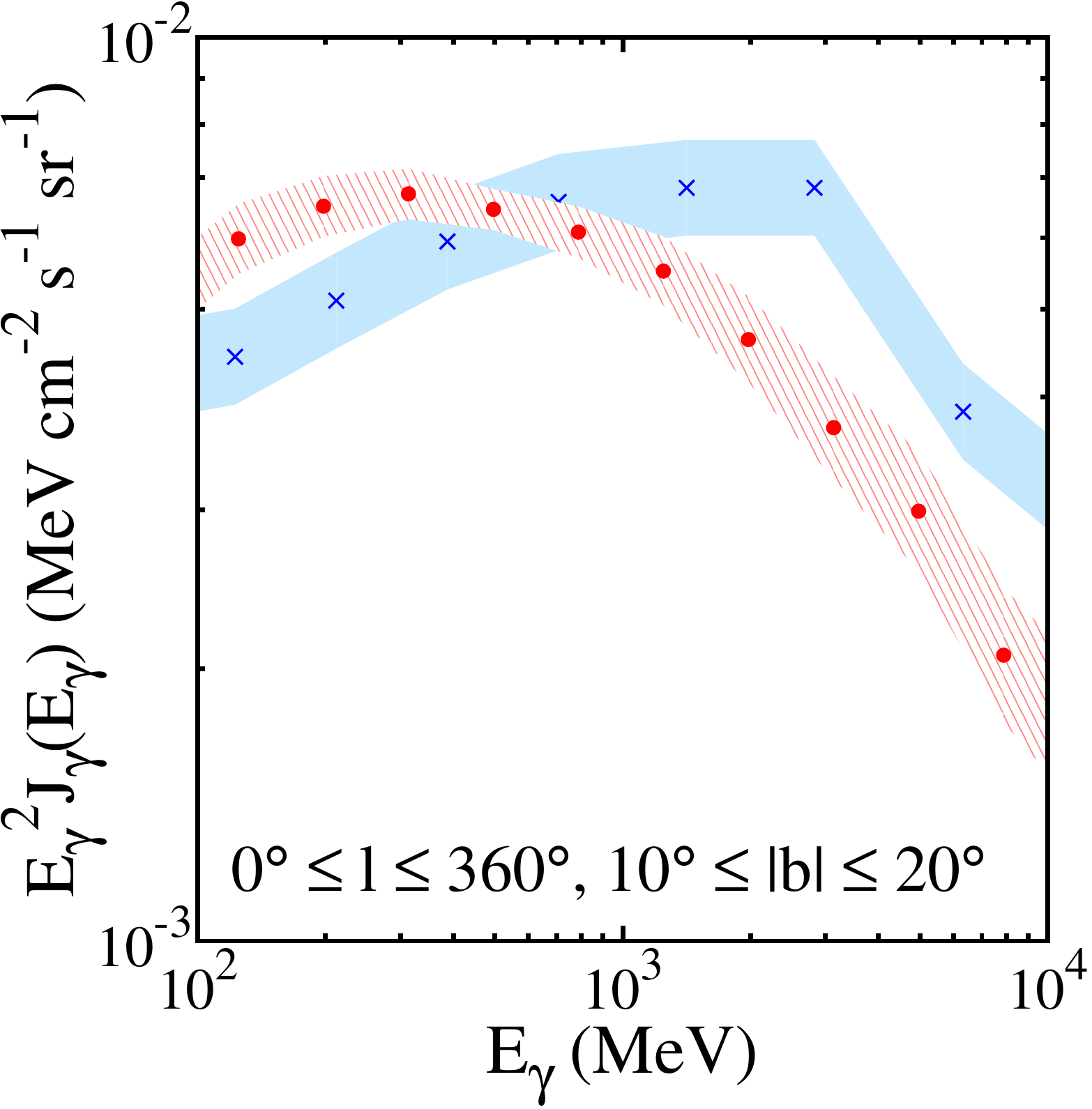}
\includegraphics[width=2.6in]{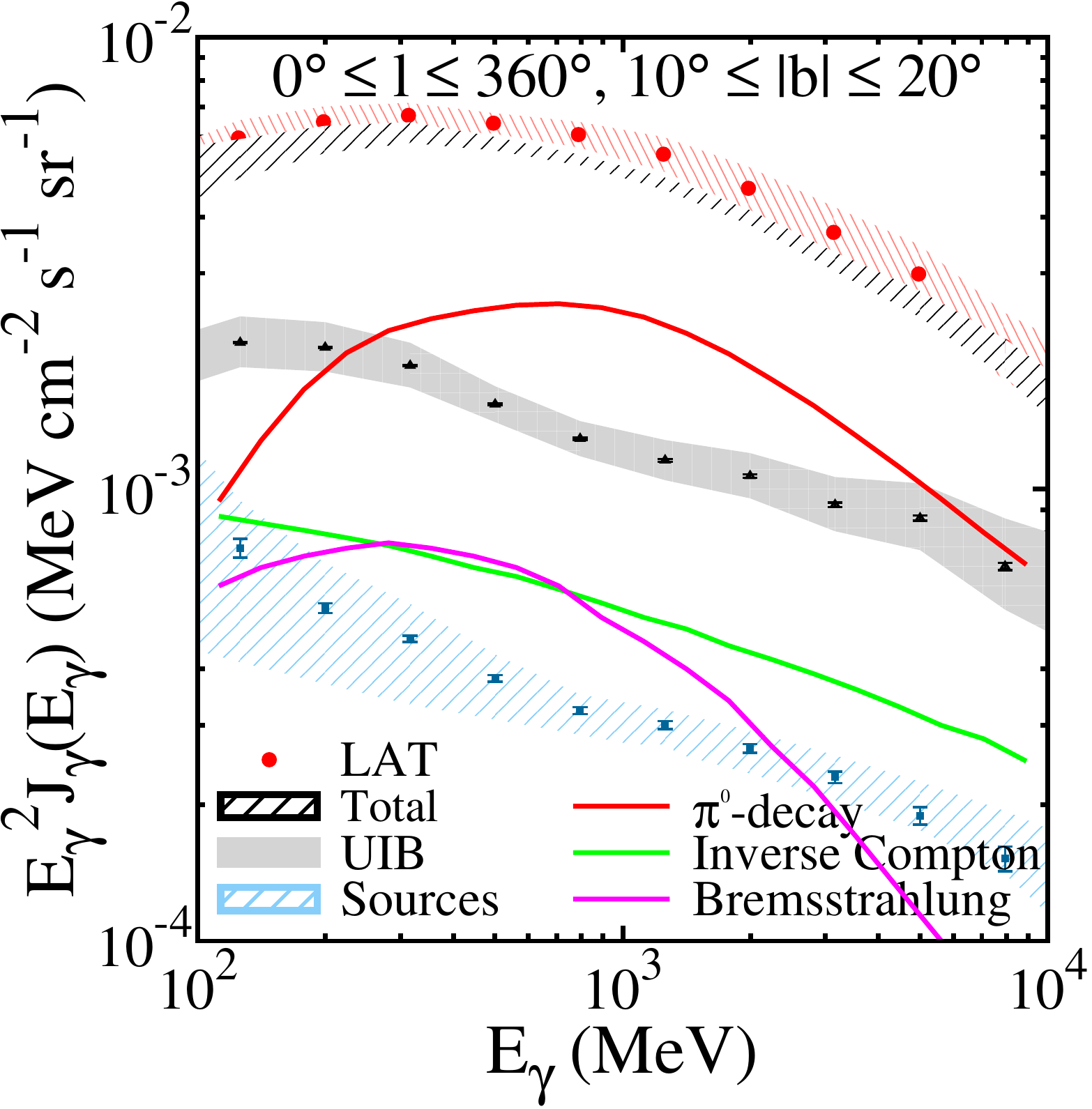}
\end{center}
\caption{
\small
Diffuse gamma-ray emission intensity from 100 MeV to 10 GeV
averaged over all Galactic
longitudes for the latitude range $10^\circ \leq |b| \leq 20^\circ$
from \citet{Abdo:2009gevexcess}.
(a): \fermilat\ (red dots) and  EGRET (blue crosses) data.
The hatched band surrounding the \fermilat\ data indicates the systematic
uncertainty in the measurement due to the uncertainty in the instrument
acceptance.
The EGRET data have the standard 13\% systematic uncertainty
\citep{Esposito1999}.
The \fermilat\ spectrum is significantly softer than the EGRET spectrum
with an integrated intensity $J_{\rm LAT}(\geq 1 \, {\rm GeV}) =
2.35\pm0.01\times10^{-6}$ cm$^{-2}$ s$^{-1}$ sr$^{-1}$
compared to the EGRET integrated intensity
$J_{\rm EGRET}(\geq 1\, {\rm GeV}) = 3.16\pm0.05\times10^{-6}$
cm$^{-2}$ s$^{-1}$ sr$^{-1}$, where the errors are statistical only.
(b): \fermilat\ data with components from a diffuse gamma-ray emission model, sources, and
unidentified isotropic background (UIB).
Model (lines): $\pi^0$-decay, red; bremsstrahlung, magenta;
inverse Compton, green.
Shaded/hatched regions: UIB (unidentified background),
gray shading; sources, blue hatching;
total (model + UIB + source), black hatching.
The UIB component was determined by fitting the data and sources with the
model held constant using the latitude range of $|b| \geq 30^\circ$, and
is a measure of the instrumental background due to charged CRs interacting
in the LAT, unresolved sources, and unaccounted diffuse Galactic gamma-ray
emission components, as well as the true diffuse extragalactic
gamma-ray background.
The spectral shape of the \fermilat\ data
is compatible with the total of the assumed model, sources, and UIB component.
No excess emission component is required.
Note that the model is {\it a priori}, based on directly measured CR
spectra, and is not
fitted to the gamma-ray data.
Thus the uncertainty in the total does not
reflect any uncertainties associated with modeling the diffuse gamma-ray
emission
\citep[which are estimated to be $\sim 20$\% for
the selected region of sky in][]{Abdo:2009gevexcess}.
}
\label{figCRgevexcess}
\end{figure}

The diffuse gamma-ray emission spectra measured by the \fermilat\ and
EGRET are shown
in Figure~\ref{figCRgevexcess}a, where no attempt has
been made to remove the contributions of gamma-ray point sources.
The \fermilat\ measured spectrum is significantly softer than that of EGRET.
The specific cause of the differences in the spectra measured
by these instruments is uncertain.
The confidence in the
\fermilat\ instrument response comes from detailed Monte Carlo
simulations
that were validated with beam tests of a calibration units, as well 
as post-launch
refinements based on actual on-orbit particle background measurements.
On-orbit studies of the Vela pulsar \citep{Abdo:2009vela1,Abdo:2010vela2}
showed similar
deviations between the \fermilat\ and EGRET spectra, so it is
unlikely that the difference is due
to differing residual particle background contamination.

Figure~\ref{figCRgevexcess}b replots the \fermilat\ data along with
the spectra of a diffuse emission model that is based on local CR
measurements, sources with $>5\sigma$ significance measured in the
first three months of data, and an unidentified isotropic background
component comprised of residual particle contamination,
unresolved sources, and unaccounted diffuse Galactic gamma-ray emission,
as well as true extragalactic diffuse emission.
It was obtained by fitting
to data at higher latitudes while holding the model constant.
Although the \fermilat\ spectral shape is consistent with the model, the overall
emission in the model is systematically low by $\sim 10-20$\%.
The diffuse gamma-ray emission model is {\it a priori}, based on
local CR measurements taken before the \fermilat\ launch and on
other information (e.g., the conversion between CO line
intensity and H$_2$ column density in the ISM, \Xco).
These uncertainties are not included in the systematic uncertainty band shown
in Figure~\ref{figCRgevexcess}b.
However, they would only shift the overall intensity higher and would not
change the spectral shape (which comes mainly from the
CR proton spectrum, measured with fairly good
accuracy to be a power law with index $\sim -2.7$).
The close correspondence between the \fermilat\ data and our expectations
based on local CR measurements and astrophysical gamma-ray emission processes
suggests that the EGRET GeV excess was the result of instrumental errors.
In fact, members of the EGRET team have called into question the calibration 
of their instrument as an explanation of the GeV excess \citep{Stecker2008}, 
but they did not identify a specific cause.

There have been other claims for the detection of excess diffuse emission
(over the standard astrophysical background) at microwave frequencies and
gamma-ray energies.
At GHz frequencies,
analysis of the \wmaplong\ (\wmap) data showed evidence for excess diffuse
emission within a $20^\circ$ radial region about the center of the Milky Way,
which was called the \wmap\ haze \citep{Finkbeiner2004}.
It is diffuse emission in excess of that expected from dust, ionized gas,
and synchrotron radiation by CR electrons,
and it has been interpreted as evidence for an additional source of
CR electrons and positrons in the inner Galaxy, possibly due to
DM annihilation or decay \citep{Hooper2007}.
Further analysis of 3-year \citep{Dobler2008,Bottino2008}
and 5-year \wmap\ data \citep{Bottino2010} also showed evidence of the excess
emission, but analysis of the 7-year \wmap\ data \citep{Gold2010} does
not show evidence for a polarized signal, casting
doubt on a synchrotron-radiation origin of the additional component.

Similarly, analysis of \fermilat\ data has led to claims
of a corresponding gamma-ray haze, described as the high-energy
counterpart to the \wmap\ haze \citep{Dobler2010} and produced by IC scattering
of electrons and positions on the ISRF.
Whether or not the \wmap\ and gamma-ray haze
are related is difficult to tell.
Furthermore, we should first
convince ourselves that the data really are in excess over the 
expected standard
astrophysical sources of diffuse emission.

The analyses of the \wmap\ and \fermilat\ data by these authors use various
fitting methods that employ templates as proxies for the
interstellar emission components comprising the
astrophysical background, e.g.,
dust maps and radio data (representing the
synchrotron emission) for the \wmap\ haze, or
gas maps and other templates to represent the various diffuse emission
components for the gamma-ray haze.
The background is removed by some procedure, and
the remaining residuals are interpreted as these excess emission components
(the \wmap\ haze or gamma-ray haze).
Sometimes the accuracy of the templates
representing the spatial morphology over the full frequency (energy) range
is debatable.
As an example,
typically for the analysis of the \wmap\ data, researchers use the 408~MHz
map of \citet{Haslam1982}, scaling with a single spectral index up to
\wmap\ frequencies.
However, studies show that there is a range of spectral indices
(from 2.3 to 3.0) between 408 and 1420 MHz \citep{Reich1988}.
So it seems
unlikely, given the large difference between the frequencies where
these surveys were made and the $\sim 20-90$ GHz \wmap\ range,
that scaling by a single index is an adequate description for this foreground
component.
The residuals extracted can be affected by these details.
For example, \citet{Bottino2008} showed how the morphology of the extracted
anomalous GHz emission from the \wmap\ data
is less extended compared to using the standard synchrotron template
when using a template based on the difference between $K$ and
$K\alpha$ \wmap\ data.

Similar care is also required with the gamma-ray data, for which
details related to the gas tracers and how the IC emission is determined
are important.
It is known that 21-cm and CO surveys are incomplete tracers
of the interstellar gas \citep[e.g.,][]{Grenier2005}.
An auxillary gas tracer, like a dust reddening map, needs to be
used to account for neutral gas not detected in 21-cm or CO
observations \citep[e.g.,][]{Abdo2ndQuad,Ackermann3rdQuad}.
Otherwise spurious sources and extended morphological features appear
as residuals simply because of our incomplete knowledge of the total gas
column density.
But correcting the gas column densities with dust maps is
only useful for regions where the dust optical depth
is not too high (outside the Galactic plane).

The IC templates are obtained from
runs of the GALPROP code (http://galprop.stanford.edu)
using a model for the ISRF that has
uncertainties, particularly in the inner Galaxy, that have not been
completely quantified \citep{Porter2005,Porter2008}.
A model is used to generate them because there is no observational
tracer for this diffuse gamma-ray emission component.
The astrophysical IC emission has a smoothly varying and extended
spatial morphology that changes depending on the assumptions
made for the underlying GALPROP run (and ISRF model) employed to calculate
the template used for an analysis.
However, DM-induced gamma-ray emission can also
have a smoothly varying and extended
morphology, because the DM provides an additional source of CR electrons
and positrons that can IC scatter the ISRF.
Disentangling the effect of incomplete knowledge of the astrophysical
emission and potential signal component(s) is difficult when
features may be introduced into the residuals due to
uncertainties in the background model templates.

Another way to search for residual emission is through direct use
of physical modeling codes, such as GALPROP.
Exploration for the origin of the \wmap\ haze and the excess gamma-ray
emission, including whether they are related, has been done using GALPROP
by several groups
\citep[e.g.][]{Lin2010,Linden2010,Mertsch2010}.
The conclusions are mixed.
Either the excesses can be attributed to uncertainties in the modeling
\citep[e.g.,][]{Linden2010,Mertsch2010}, or other components are
required that could have either astrophysical or DM origins
\citep[e.g.,][]{Lin2010,Meng2010}.
However, these models involve
a decomposition of \hi\ and CO gas maps into
galactocentric rings to represent
the distribution of interstellar gas, together with other details such
as assumed distributions for the CR sources, uncertainties
associated with the ISRF (as discussed above), and unresolved
point source populations \citep[see, e.g.,][]{Strong2007popsyn}.
For the \wmap\ haze the uncertainties in the Galactic magnetic field
remain important.

With physical models parameter scans can be made to
quantify the systematic uncertainties.
An example of an analysis involving a subset of model parameters using
the gamma-ray data and GALPROP code is
the \fermilat\ team analysis of the isotropic gamma-ray background
\citep[see table~1 in][]{Abdo:2010iso}.
It was restricted to Galactic latitudes $|b| \geq 10^\circ$ and investigated
the systematic uncertainties associated with the local gas distribution and
IC contribution when modeling the Galactic diffuse gamma-ray emission as
a foreground to the isotropic gamma-ray emission.
The residuals had characteristics similar to the
gamma-ray haze found by \citet{Dobler2010} (for the regions of sky common
to both analyses), but the morphology changed
non-linearly under variation of the model parameters, such that the 
effects of individual parameters were
not easily separable.
Thus the residuals and their morphological changes were
ascribed to systematic uncertainties in the foreground modeling for the
\citet{Abdo:2010iso} analysis,
without identification of any single model parameter as being the cause.

Recent work with Bayesian methods and the GALPROP code has focused on
obtaining constraints for the parameters of
diffusive-reacceleration CR propagation models using CR nuclei data
\citep{Trotta2011}.
The results are consistent with earlier less rigorous analyses, but 
\citet{Trotta2011} also provide statistically well-motivated uncertainties
for the model parameters.
In the future
such studies can be extended to full parameter scans, including the diffuse
emission and other CR data, such as antiprotons and positrons.
This would enable a better understanding, on a firmer statistical foundation,
of the uncertainties involved when modeling
CRs and diffuse Galactic emission.
Meanwhile, although
the \wmap\ haze and gamma-ray excesses are interesting and motivate
alternative physical explanations of their origin,
we think that it is premature to draw conclusions on
whether DM is needed or more prosaic explanations are sufficient.

Although the \fermilat\ data convincingly rule out the EGRET large GeV 
excess, there
still is much interest in looking for a smaller excess in diffuse 
gamma-ray emission from the Galactic halo.
Unfortunately, the same issues as discussed at length above in relation to the
\wmap\ and gamma-ray haze also affect the halo analyses.
Whether a tool like GALPROP is used to predict the background emission or
templates are used, in either case it is difficult to quantify the modeling 
uncertainties.
These issues have thus far prevented the \fermilat\ collaboration from
publishing limits on DM annihilation in the Galactic halo based on their
all-sky diffuse emission data.
The same issues are also critical for interpretation
of gamma-ray data for
the very innermost regions toward the GC, because the unfolding
in this region of the sky depends strongly
on the uncertainties associated with distributions of the
sources of diffuse emission, not to mention a potentially large
number of point sources.

In summary, the diffuse emission from the Galaxy provides the
highest statistical power for searching for DM signals.
But the challenges in
understanding the astrophysical backgrounds are also high and require
significant further work before strong conclusions can be made.

\subsection{Extragalactic Diffuse Emission}
\label{sec:extragalacticdiffuseemission}

\begin{figure}
\begin{center}
\includegraphics[height=2.0in]{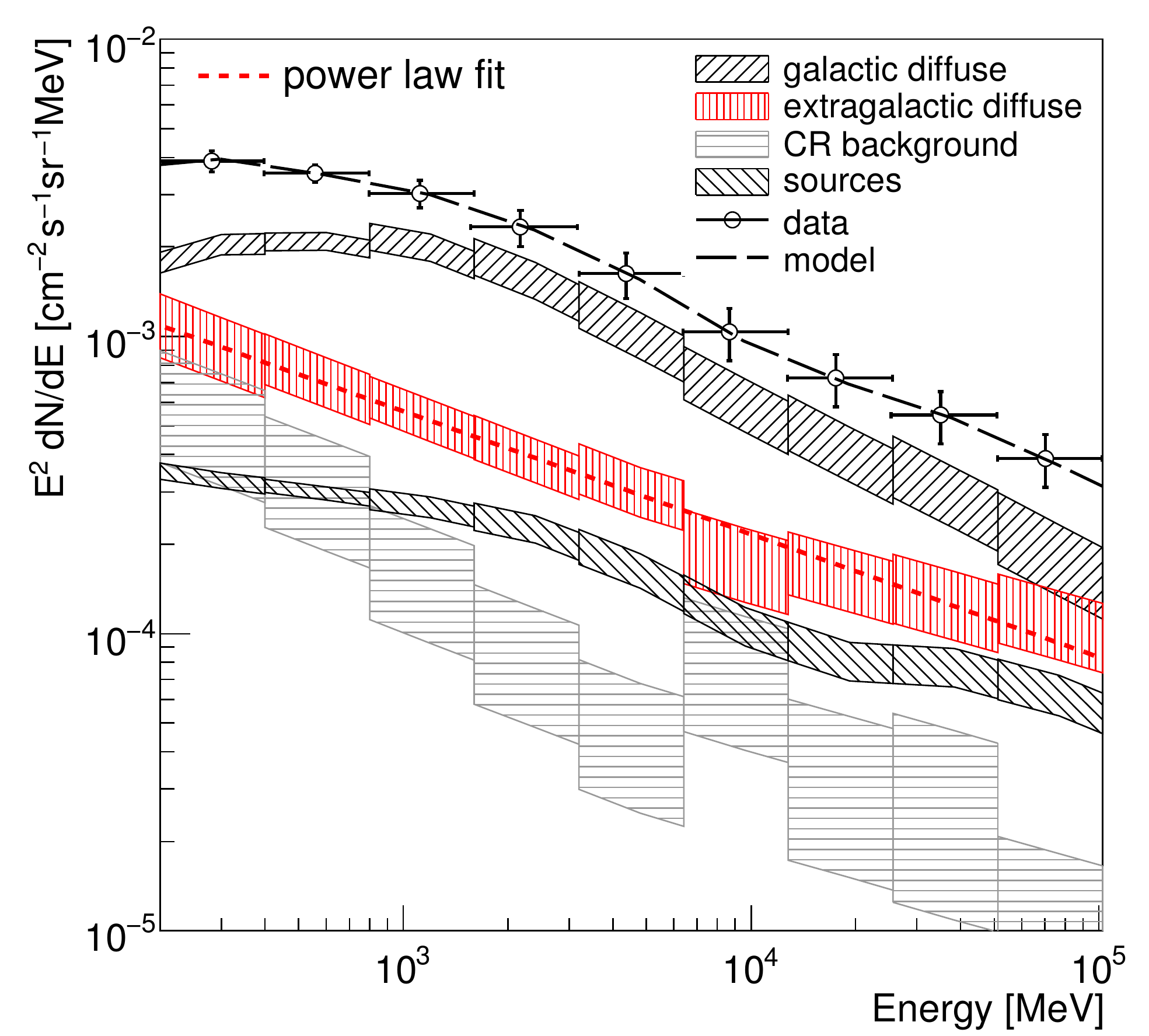}
\includegraphics[height=2.15in]{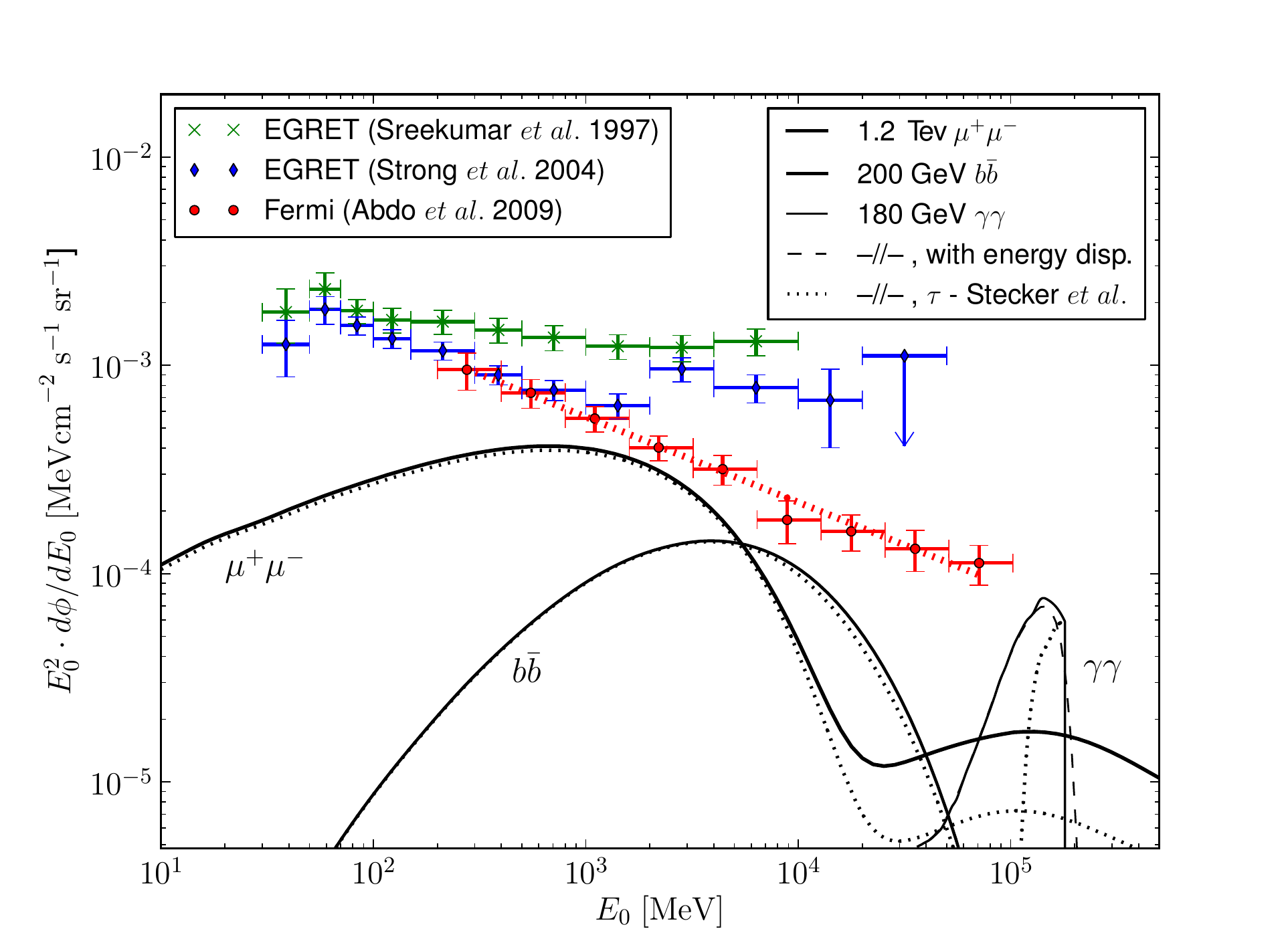}
\end{center}
\caption{\small
(a) \fermilat\ measured gamma-ray intensity with fit results
for the energy range 100 MeV to 100 GeV and averaged over
all Galactic longitudes for $|b| \geq 10^\circ$.
Fit results by component are given in table~1 of \cite{Abdo:2010iso}.
Note that the statistical errors for the \fermilat\ data are smaller than
the data symbols, and the systematic uncertainties from the instrument
effective area dominate the LAT data for the energy range
shown in the figure.
The \fermilat-derived spectrum for the extragalactic gamma-ray background 
(shown as the red shaded region) 
is compatible with a simple power law with index $\gamma = 2.41 \pm 0.05$ 
and intensity
$I(>100\,{\rm MeV}) = (1.03 \pm 0.17)
\times 10^{-5}$~cm$^{-2}$~s$^{-1}$~sr$^{-1}$.
(b) Extragalactic gamma-ray background spectra
derived from \fermilat\ data \citep{Abdo:2010iso}
and EGRET data [taken from table~1 of \citet{Sreekumar1998} and table~3
of \citet{SMR2004egb}], together with three potential 
types of gamma-ray spectra induced by dark matter (DM) considered 
in the analysis by \citet{Abdo:2010cosmo} (these spectra are 
good approximations for the likely signals from the different final states 
discussed in section~\ref{sec:detectability} and summarized in 
table~\ref{Tab:finalstates}).
The overall normalization of the DM spectra are obtained 
assuming a particular extrapolation method for the
structure and sub-structure contribution to the DM signal from the
Millenium-II numerical 
calculation \citep{MilleniumII} [the most conservative scenario
considered for the evolution of structure 
in \citet{Abdo:2010cosmo}].
The limits for the annihilation cross sections for the different models 
are $1.2\times 10^{-23}$ cm$^{3}$ s$^{-1}$ (for a 1.2~TeV WIMP 
annihilating to $\mu^+ \mu^-$),
$\langle \sigma v\rangle= 5\times 10^{-25}$ cm$^{3}$ s$^{-1}$ (for a 200~GeV 
WIMP annihilating to $b {\bar b}$), and $2.5\times 10^{-26}$ cm$^{3}$ s$^{-1}$ 
(for 180~GeV WIMP annihilating to $\gamma \gamma$).
The solid lines in the figure correspond to the various 
DM spectra, including the effects of gamma-gamma absorption on the 
extragalactic background light (EBL) using the model of \citet{Gilmore2009}.
The dotted lines show the DM spectra, including the effects of 
gamma-gamma absorption on the EBL using the
model of \citet{Stecker2006}.
The dashed lines are the DM model spectra, including the finite energy 
resolution of the \fermilat\ for the \citet{Gilmore2009} EBL model. 
For the $b{\bar b}$ and $\mu^+\mu^-$ final states, the limits are 
$10-1000$ times higher than the cross section
given by Equation~\ref{eqn:standard_cross_section} [limits for the
DM models with relatively 
strong gamma-ray lines are below the current sensitivity of the \fermilat\ 
\citep[see][for details]{Abdo:2010cosmo}].
}
\label{figGammaIsotropic}
\end{figure}

The diffuse Galactic emission presents a strong foreground signal to the
much fainter diffuse extra-galactic emission, which is
often referred to as the extra-galactic gamma-ray background (EGB) and 
generally assumed to have an isotropic or nearly isotropic
distribution on the sky.
The EGB was first detected against the much brighter diffuse Galactic emission
by the {\it Small Astronomy Satellite 2} (SAS-2)
satellite \citep{Fichtel:1975} and later confirmed
by analysis of EGRET data \citep{Sreekumar1998}.
The EGB is composed of contributions
from unresolved extragalactic sources as well as truly diffuse
emission processes,
such as possible signatures of large-scale
structure formation \citep{Waxman:2000}, emission produced by the
interactions of
ultra-high-energy CRs with relic photons \citep{Kalashev:2009}, the
annihilation or decay
of DM, and many other
processes \citep[e.g.,][and references therein]{Dermer2007}.
However, diffuse gamma-ray emission from IC scattering
of CR electrons in an extended Galactic halo can also produce a very 
smoothly varying (close to isotropic) 
spatial distribution that could also be ascribed to 
the EGB if the 
size of the halo is large enough (that is, $\sim 25$ kpc) \citep{Keshet:2004}.
Because the EGB is extracted using gamma-ray data at high Galactic 
latitudes it is difficult to distinguish between foreground contaminants
like an extended halo and the true EGB, so there is no 
assurance that a detected isotropic component has an extragalactic origin.
Nevertheless, we shall continue to use the term EGB when discussing it.

The latest observational contribution to this subject is
the \fermilat\ measurement of the spectrum of
isotropic diffuse gamma radiation from 200~MeV to
100~GeV \citep{Abdo:2010iso}.
The biggest challenge, and the largest source of systematic uncertainty, in a
measurement of the EGB is the subtraction
of the various foregrounds.
Most important are the diffuse Galactic emission,
the contribution from resolved sources, and
the instrumental background from misclassified CRs.
In the \citet{Abdo:2010iso} analysis
the misidentified CRs were suppressed by applying very stringent
event selection criteria, albeit at the expense of efficiency.
The diffuse Galactic gamma-ray emission was modeled using the GALPROP code
(with particular attention given to characterizing the dominant sources
of systematic uncertainties in the foreground modeling; see 
Section~\ref{sec:diffuseemission}). 
The isotropic background was then found using a simultaneous fit of the 
diffuse Galactic gamma-ray emission from the modeling, 
resolved sources from the internal \fermilat\ 9-month source
list (using the individual localizations but leaving the fluxes
in each energy bin to be separately fitted for each source), and 
a model for the solar IC gamma-ray emission.
The fitted isotropic component contained the contributions by 
misidentified CRs that still passed the event selection and the EGB.
The residual particle backgrounds were isotropic over the data taking 
period for the \citet{Abdo:2010iso} analysis, 
estimated using the instrument simulation, and then subtracted from the 
fitted isotropic component to obtain the EGB.
The derived EGB spectrum is shown in Figure~\ref{figGammaIsotropic}a.
It is a featureless power law,
significantly softer than the one obtained from EGRET observations
\citep{Sreekumar1998}, as can be seen in Figure~\ref{figGammaIsotropic}b.
Also, the spectrum does not show a feature at $\gtrsim 2$ GeV found in a
reanalysis of the EGRET data with an updated diffuse gamma-ray emission 
model based on GALPROP \citep{SMR2004egb}.

Using the \fermilat-derived EGB, \cite{Abdo:2010cosmo} set upper limits on the
gamma-ray flux from cosmological annihilation of DM.
Figure~\ref{figGammaIsotropic}b shows the gamma-ray spectra for
representative particle physics models used in the analysis.
Several models of varying degree of optimism were considered for
the cosmological evolution of structure.
WIMP annihilation is
very sensitive to the amount of substructure clumping, which enhances
the density-squared in the integrand of
$J(\psi)$ (Equation~\ref{eqn:astrofactor}).
The most conservative structure evolution model 
assumed only the results of a large numerical
many-body calculation, and the results for this model are shown in 
Figure~\ref{figGammaIsotropic}b for three representative annihilation
final states ($\mu^+ \mu^-$, $b {\bar b}$, and $\gamma \gamma$).
The more optimistic models treated the substructure contributions by
analytically extrapolating to much smaller scales.
Different models for the extragalactic
background light were also
considered to gauge the effect of gamma-gamma absorption.
The most conservative limits for the DM annihilation cross section
were obtained assuming that all of the observed emission was
due to DM.
The limits for the DM annihilation cross sections and masses were:
$1.2\times 10^{-23}$ cm$^{3}$ s$^{-1}$ 
(for a 1.2~TeV WIMP annihilating to $\mu^+ \mu^-$),
$\langle \sigma v\rangle= 5\times 10^{-25}$ cm$^{3}$ s$^{-1}$ (for a 
200~GeV WIMP annihilating to $b {\bar b}$), and 
$2.5\times 10^{-26}$ cm$^{3}$ s$^{-1}$ (for a 180~GeV WIMP annihilating to 
$\gamma \gamma$).
Less conservative constraints were derived by first subtracting
simple models for the contributions from unresolved star-forming 
galaxies and blazars [which were modeled assuming power-law spectra 
with indices $-2.7$ (star-forming galaxies, motivated by 
theoretical calculations) and $-2.4$ (blazars, from observations)].
[Note, the contributions by the different astrophysical source classes 
to the EGB is not well known. Recent work by the \fermilat\ 
collaboration \citep{Abdo:EGBOrigin} has derived a fraction $<40$\%
for the contribution by blazars to the EGB. The contributions by other 
source classes are so far undetermined.]  
Under these assumptions, very good constraints for 
the DM annihilation cross section 
were obtained in the most optimistic scenarios, even excluding
the expected thermal WIMP cross
section (Equation~\ref{eqn:standard_cross_section}) up to 1~TeV.
However, the true upper limits that are based on 
the conservative assumptions were a factor
$10-1000$ times higher and
realistically 
do not exclude any of the MSSM parameter space for the annihilation
process and cross section.

\subsection{Gamma-Ray Line Searches}

The cleanest and most convincing DM
signal that could be measured would be a monochromatic gamma-ray line
on top of the continuum background spectrum.
A gamma-ray line would result from decays or annihilations to
two-body final states:
$\chi(\chi) \rightarrow \gamma X$, where $X$ can be another
photon, a $Z$ boson, a Higgs boson, or a non-SM
particle \citep[e.g.,][]{Bouquet1989}.
However, DM models predict branching fractions into such                        
states that are typically $\sim 10^{-4}-10^{-1}$ compared to the total          
annihilation or decay rate, placing them below the                              
flux sensitivity of any existing instrument.
A search for gamma-ray lines in the range of $0.1-10$~GeV was done using EGRET
in a $10^\circ \times 10^\circ$ region about the
GC with a null result \citep{Pullen2007}.
Most recently, the
\fermilat\ team used the first year of survey data
to search for gamma-ray lines in
the energy range of $30-200$~GeV \citep{Abdo:2010lines}.
The analysis used data from two regions:
all the sky but the Galactic plane (that is, $|b| \geq 10^\circ$ to
avoid the strong diffuse emission from CRs interacting with
interstellar gas) and also
a $20^\circ \times 20^\circ$ region centered on the GC.
Even though source and diffuse emissions are the strongest and most
complex around the GC, this region was included in the analysis
because the DM concentration also peaks in there, significantly
increasing the potential signal.
Known point sources were removed by cutting out a $0.2^\circ$ circular
region around each, except within $1^\circ$ of the GC, in order not
to remove this entire region.
The resulting spectrum showed no evidence of a line.
Flux limits were set at 10~GeV intervals, starting at 30~GeV.
For each energy the spectrum was fit to a line shape corresponding to
the predicted instrument response and centered on that energy, together with
a polynomial background.
For each of three assumed Galactic distributions of DM,
the flux limits were translated into cross sections for annihilation to
$\gamma\gamma$ or $\gamma Z_0$ as well as lifetimes for
decay to the same final states.
See table~1 of \cite{Abdo:2010lines}.
The annihilation cross section upper limits were of
order
$\langle \sigma_{\gamma\gamma} v \rangle \sim 10^{-27}\,{\rm cm^3 \, s^{-1}}$,
and the lifetime lower limits were of order $\tau_{\gamma\gamma}\sim 10^{29}$~s.
Unfortunately, these limits do not significantly constrain                      
the parameter space of DM models for a typical thermal WIMP because of          
the small branching fractions typically expected for these                      
annihilation or decay modes (an increase in sensitivity of $\sim 100-1000$      
over the current measurements would be required to constrain canonical          
thermal WIMP models).
Nevertheless, they do disfavor or rule out particular
models, such as some
that have been constructed to explain the \pamela\ rising positron abundance,
some that rely on non-thermal production of 
WIMPs (Equation~\ref{eqn:standard_cross_section}), and
models with a thermal cross section but constructed to have a large branching
fraction to a gamma-ray line \citep[e.g.,][]{Goodman2010}.

\subsection{Gamma Rays from Dwarf Spheroidal Galaxies and Satellites}
\label{sec:dwarfgalaxies}

\begin{figure}
\begin{center}
\includegraphics[height=3.3in]{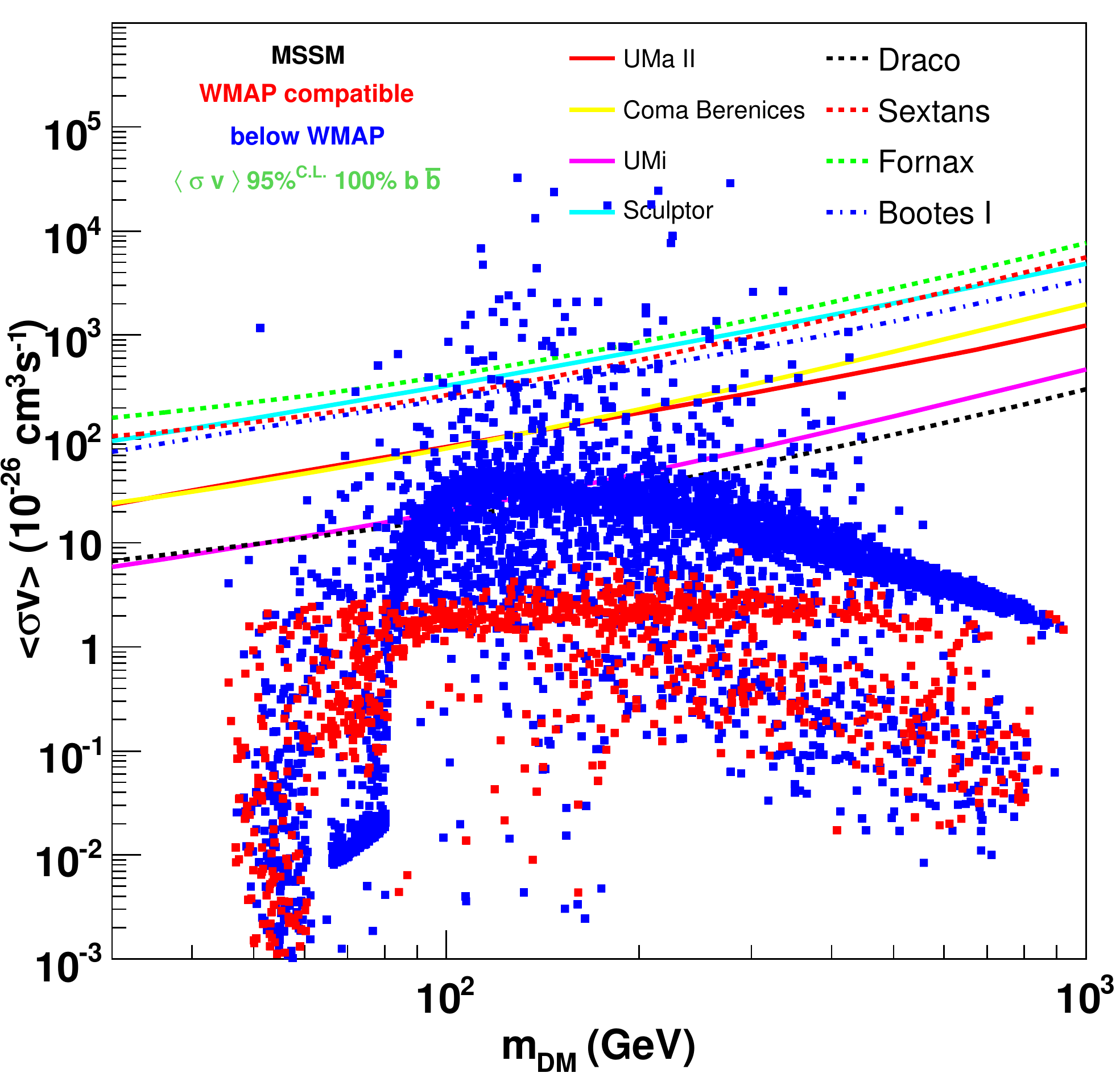}
\caption{\fermilat\ 95\% upper limits for WIMP annihilation to $b\bar b$ in
selected dSph galaxies in the ($m_{{\rm wimp}}$,$<\sigma v >$) 
plane, from \cite{Abdo:2010ex}.
The points are derived from a scan over 7 parameters of the
Minimal Supersymmetric Standard Model.
The red points are the most interesting, as they correspond to
a thermal relic density compatible with WMAP data.
Thus they roughly correspond to a WIMP annihilation
cross section as given by Equation~\ref{eqn:standard_cross_section}, but
complex interactions involving multiple supersymmetric partners in the early 
Universe
can change the prediction significantly, as indicated.
The blue points represent higher cross sections, and correspondingly lower
thermal relic densities, but assume that additional non-thermal production
mechanisms contribute to WIMP production, such that WIMPs still comprise 
all of the dark matter.
}
\label{figFermiDwarfbbbar}
\end{center}
\end{figure}

\begin{figure}
\begin{center}
\includegraphics[height=2.2in]{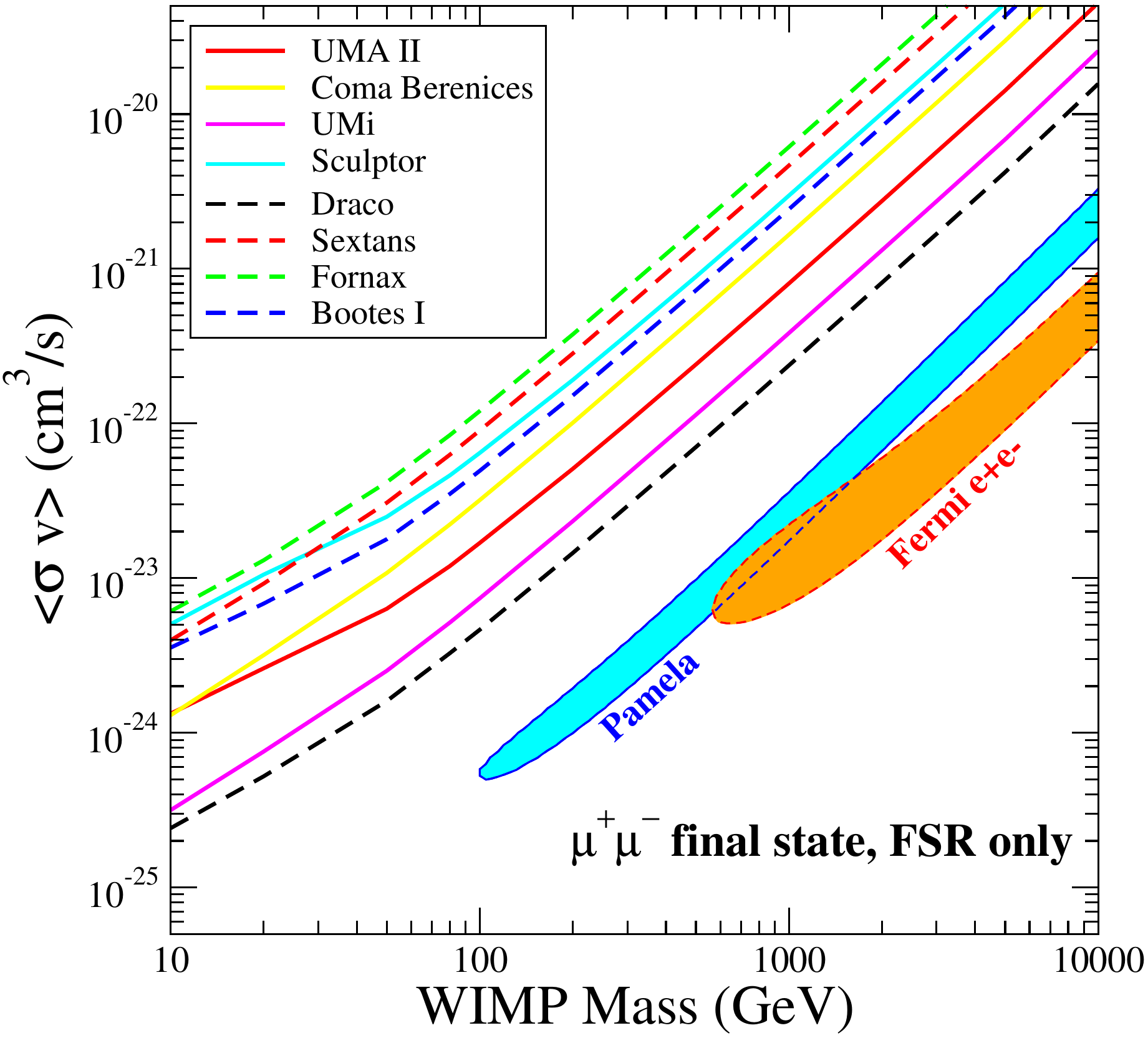}
\includegraphics[height=2.2in]{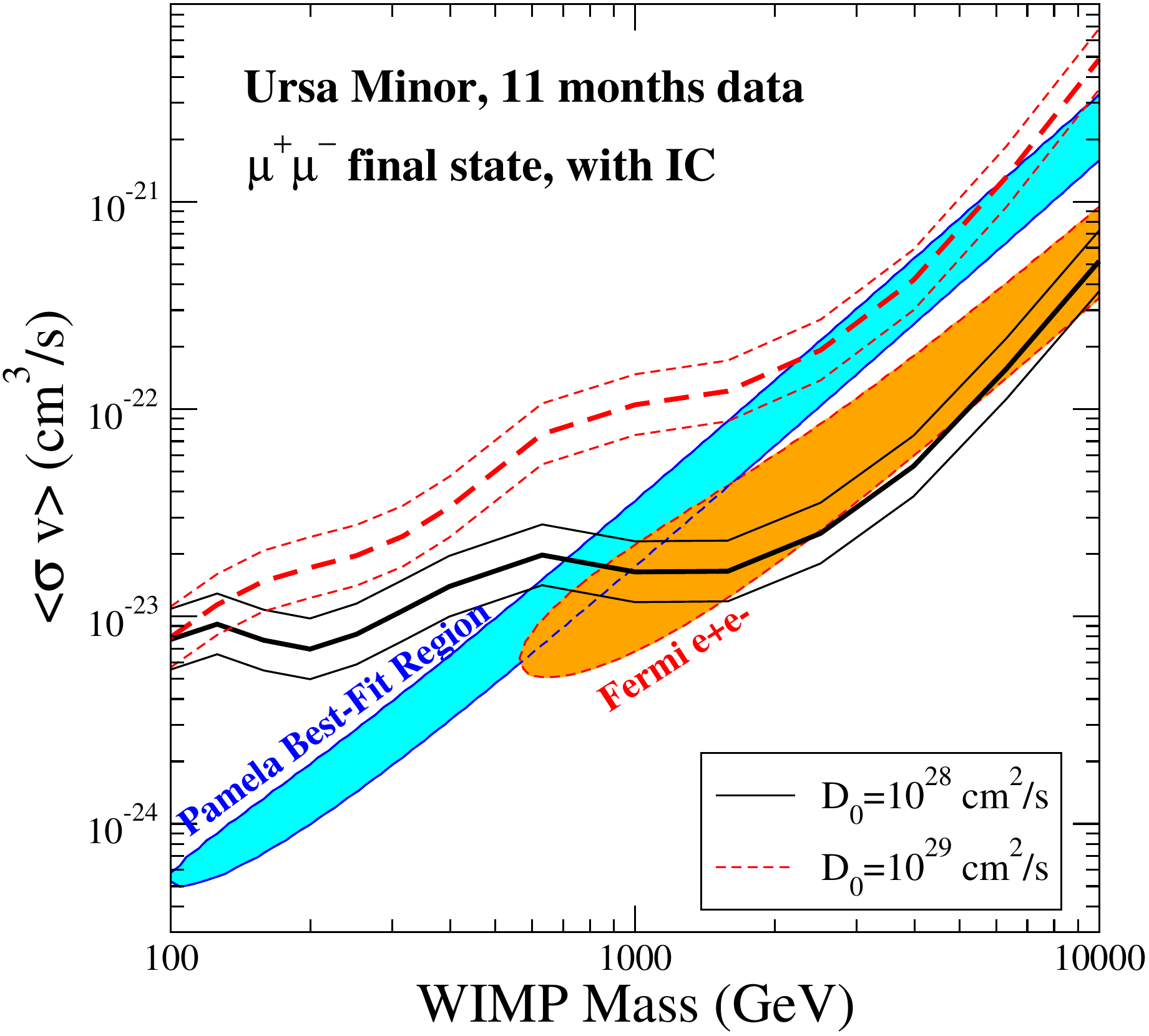}
\caption{\fermilat\ upper limits for dark matter 
annihilation to $\mu^+\mu^-$ in
dSphs \citep{Abdo:2010ex}, compared to models that fit well either the
\pamela\ measurement of the positron fraction or the \fermilat\ measured 
total electron spectrum.
The left panel shows the constraints considering
gamma-ray emission from final state radiation only.
The right panel shows the constraints for the Ursa Minor
dwarf including both
final state radiation and emission from inverse Compton 
scattering of the CMB by the
positron and electron muon-decay products, for two different
assumptions for the cosmic ray diffusion coefficient.
The bands indicate the effect of uncertainties in
the Ursa Minor dark matter density profile.
Including the inverse Compton 
contribution improves the upper limit, but at the expense of
using a model dependent upon an unconstrained diffusion coefficient.}
\label{figFermiDwarfLepton2}
\end{center}
\end{figure}

\begin{figure}
\begin{center}
\includegraphics[width = 4.5 in]{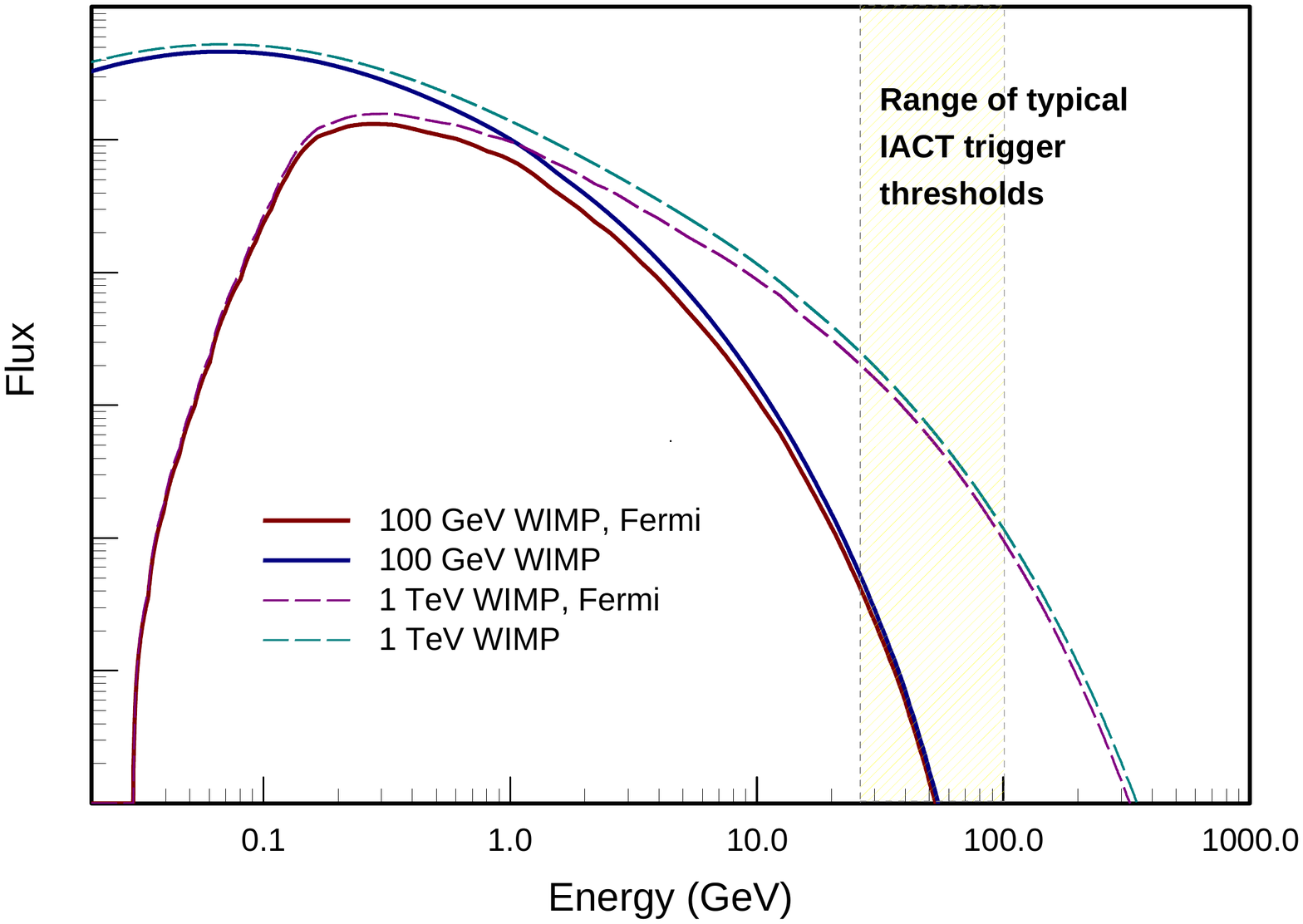}
\caption{The spectrum of detected gamma rays from dark matter annihilation 
to $b\bar b$, as predicted by DarkSUSY for WIMPs of 100~GeV and 1~TeV mass.
In each case the rate is compared between an ideal 1~m$^2$ instrument and
the \fermilat, taking into account its effective area vs.\ energy for
``diffuse-class'' events \citep{Atwood09}.
The range of {\em trigger} thresholds for ground-based \iactlong{s} 
is also shown, indicating that their sensitivity to this final state, 
even with their $> 10,000\,{\rm m^2}$ effective areas (but higher 
background and shorter observing time),
is relatively small until the WIMP mass exceeds a TeV.}
\label{figDMSpectrum}
\end{center}
\end{figure}

\begin{figure}
\begin{center}
\includegraphics[width = 4.5 in]{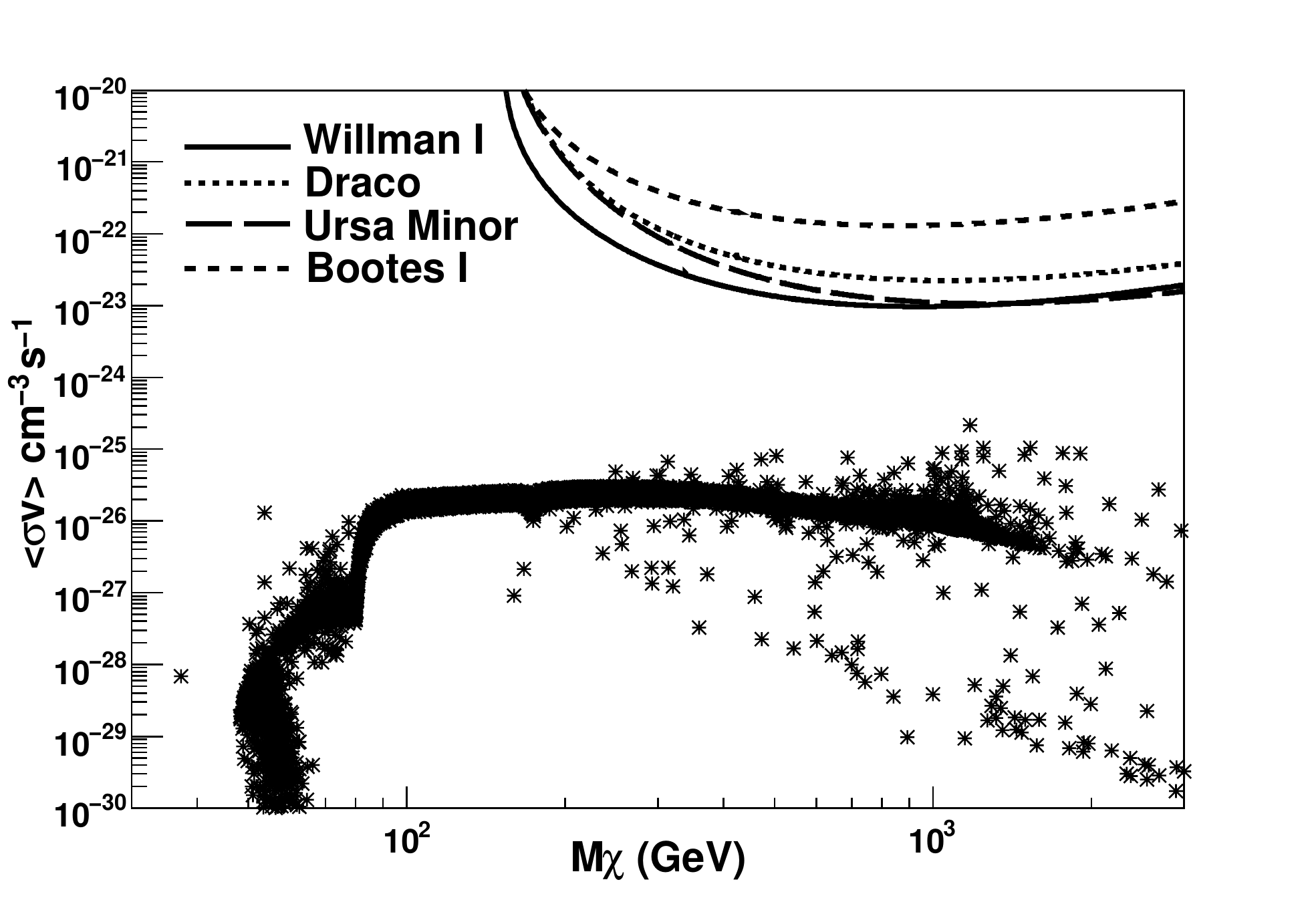}
\caption{DM annihilation upper limits from
observations of four dSphs by \veritas, from \cite{Acciari:2010pj}.
The $\Phi_s(E)$ from Equation~\ref{eqn:PPfactor} is taken to be a sum over
final states with branching fractions from a generic model, but it is
comparable to assuming a 100\% $b\bar b$ final state.
Black asterisks represent minimal supersymmetric model 
predictions for neutralino WIMPs with thermal relic abundance corresponding 
to the inferred cosmological dark matter density.}
\label{figVeritas_dSph}
\end{center}
\end{figure}

In the standard cosmological model the formation of structure, and of
galaxies in particular, is dominated by cold DM (CDM) and
proceeds hierarchically, with small structures merging over time
into ever larger galaxies and clusters \citep{Blumenthal84}.
Numerical studies of this process predict that the Milky Way should be
surrounded by a swarm of smaller structures, including gravitationally
unbound streams as well as bound subhalos \citep{Zemp09},
many of which would be expected to contain baryonic matter in the form of
stars.
A couple of dozen Milky Way satellite galaxies are
known \citep{Bullock10}, half of them discovered since 2004 by the Sloan
Digital Sky Survey \citep[SDSS;][]{York00}.
But that number is small compared to the hundreds predicted by
simulations \citep{Klypin99,Tollerud08}.
The difference is partly due to the fact that the SDSS has surveyed
only $\sim 20$\% of the full sky, and even in the surveyed regions
dim satellites are difficult to detect against the Galactic foreground.
Also, SDSS is biased toward seeing the nearer satellites, and those are the most likely to be
tidally disrupted.
The remaining difference might be understood in terms of truly dark
satellites, that is, those without sufficient baryonic matter to form stars.
This is supported by studies showing that even many of the
known satellites have very low stellar contents compared to their
DM masses, the more so for the smaller objects \citep{Simon07}.
Several physical processes could prevent baryons from accumulating in small, shallow
potential wells, such as heating from early reionization of the universe \citep{Somerville2002}.
Therefore, ongoing searches for DM consider both the known satellites
and unknown dark satellites.
The former clearly are easier to treat, because their locations are known,
as are their approximate mass-to-light ratios.

Dwarf spheroidal (dSph) galaxies (a subclass of the dwarf elliptical
galaxies) are the most attractive candidate subhalo objects for
DM searches, because these low-luminosity structures are typically
found by line-of-sight stellar velocity dispersion measurements to have
very high mass-to-light ratios \citep{Bullock10} and thus abundant DM.
They probably also have low astrophysical gamma-ray backgrounds, because
they contain almost no gas or dust, no star-forming regions, and probably
few millisecond pulsars, compared with globular clusters, which are
rich in millisecond pulsars.
Because dSphs are small and localized, \iact{} experiments, as well
as the \fermilat, can contribute to the search for gamma rays
produced by DM annihilation or decay.

The stellar measurements indicate that the known dSphs all have
roughly the same mass within their central 300~pc, independent
of their luminosities \citep{Strigari08},
in which case the best targets for indirect detection of DM should be
those that are closest to Earth but not obscured by the Galactic disk.
Thus far, no statistically significant gamma-ray excess has been detected
from any dSph location by any of the gamma-ray telescopes.
The \fermilat\ collaboration has published flux upper limits for 14 dSphs
based on the first 11 months of data \citep{Abdo:2010ex}.
The limits are presented for several assumptions on the source
spectra: power laws of five spectral indices, $\Gamma$, ranging
from 1 to 2.4 versus DM annihilation for WIMP masses between 10~GeV
and 1~TeV.
Several final states are considered for the DM annihilation,
including $b\bar{b}$, $W^+W^-$, $\tau^+\tau^-$, and $\mu^+\mu^-$.
For $\Gamma=2$ the typical integral flux upper limit above 100~MeV for a single dSph location is a
few times $10^{-9}$ photons cm$^{-2}$ s$^{-1}$.

To interpret the flux limits as constraints on DM annihilation or decay
requires knowledge of the amount of DM in each dSph.
In addition, DM annihilation is highly sensitive to the radial distribution and central concentration of
the WIMPs, as it is proportional to the square of the
density (Equation~\ref{eqn:dwarfFlux}).
The stellar line-of-sight velocity dispersion measurements are effective
in constraining the dSph mass and, to some degree, the DM distribution
within; but they cannot measure the
shape of the central DM cusp, let alone small-scale clumping, to a degree that
would allow a completely model-independent assessment of
the geometrical part of Equation~\ref{eqn:dwarfFlux}.
Therefore, the \fermilat\ analysis assumes a smooth Navarro-Frenk-White
profile \citep{Navarro96} out to the dSph tidal
radius $r_t$ (see \cite{Read05}), conservatively with no boost from
substructure or possible long-range attractive interactions between
WIMPs \citep[Sommerfeld enhancement, e.g.,][]{ArkaniHamed:2008qn}.
For each dSph, the Navarro-Frenk-White
characteristic density $\rho_s$ and scale
radius $r_s$ must be evaluated from the stellar data via an involved
procedure \citep{Abdo:2010ex} that relies on prior probabilities
for $r_s$ and $\rho_s$ derived from $\Lambda$CDM simulations
\citep{Diemand07,Springel08} as well as prior probabilities for the stellar
mass-to-light ratio and the velocity dispersion anisotropy.
In the case that photons come from DM decay rather than annihilation, the
astrophysical model is less critical, as the flux is then simply proportional
to the amount of DM mass along the line of sight and within the instrument PSF.
However, \citet{Abdo:2010ex} do not provide limits for decay lifetimes.

In \citet{Abdo:2010ex} only a select 8 of the 14 dSphs are used to set
limits for the DM annihilation cross section.
Results are shown versus WIMP mass for each of the 8 in
Figure~\ref{figFermiDwarfbbbar}, where each red or blue point
represents the
theoretical prediction of the MSSM for a given set of model parameters.
The red points have a thermal relic abundance corresponding to the
inferred cosmological DM density, thus forming a band along the standard
cross section of Equation~\ref{eqn:standard_cross_section}, where the deviations
from that standard follow from model details such as
the spectrum of supersymmetric partners that ``co-annihilate'' in the early 
Universe.
That is, before freeze-out in the early Universe there may have been 
multiple supersymmetric particle types 
of similar mass interacting, not just the lightest supersymmetric partner 
that is the DM WIMP, so the relic
abundance depends on more than just the annihilation cross section 
of the WIMP.
The blue points represent models with higher annihilation 
cross sections, corresponding to lower thermal relic densities.
They still assume that WIMPs comprise all of the dark matter and thus 
rely on esoteric models
in which there are additional non-thermal production processes.
Similar plots are provided by \cite{Abdo:2010ex} for other particle physics
models, including Kaluza-Klein universal extra dimensions and mSUGRA, but
the main point here is
that with $\sim 10-20$\% of the eventual complete \fermilat\ data set in
hand, the limits from individual dSphs are still a factor of 10 or more above
the most interesting parameter space pointed to
by Equation~\ref{eqn:standard_cross_section}.

The dSph Segue~1 is not included in the analysis by \cite{Abdo:2010ex} 
because of
controversy over whether it is a dSph or merely a star cluster stripped
from the Sagittarius galaxy \citep{NiedersteOstholt09}.
A more recent publication makes a strong case, based on recent
spectroscopic observations, for it to be a dSph and, in fact, the most
DM-dominated galaxy known \citep{Simon10}.
It is arguably the best target for DM searches, due to its
proximity (only 25~kpc from the Sun) and high Galactic
latitude, as well as its high DM mass.
The \fermilat\ collaboration has not yet presented DM limits from Segue~1,
but analyses based on flux limits from 9 months of data have
been published by 
subsets of collaboration members \citep{Scott09,Essig10}.
Besides including more dwarfs such as Segue~1, the results
of \cite{Abdo:2010ex} will be strengthened by an ongoing combined analysis
of all dSphs.

For $\mu^+\mu^-$ final states
the only direct gamma-ray signal comes from
the hard $E^{-1}$ photon
spectrum of final-state radiation ($\tau^+\tau^-$ final states are similar 
but do also produce some
photons from $\pi^0$ decay).
However, additional lower-energy photons are generated by IC scattering of
the CMB (which is the dominant
radiation field in dSphs due to the paucity of stars and dust)
by high-energy electrons and positrons from $\mu^\pm$ decay as they 
propagate through the galaxy.
The \iact{s}, with their
high energy thresholds, would not see that secondary production, but 
for the \fermilat\ it
can comprise a significant fraction of the signal.
This complicates the analysis by introducing a dependence on
CR propagation in the dSph, which is not necessarily well described by
the same models used to describe CR propagation in the Milky Way.
Nevertheless, the analysis employs the usual diffusion-loss equation,
solved in spherical symmetry with free-escape boundary conditions.
The results depend on the diffusion coefficient, which is not constrained by
any existing data but can only be assumed to be in the neighborhood of the value
relevant to the Milky Way.
A larger coefficient results in more of the photon signal being produced
outside of the vicinity of the dSph covered by the telescope PSF, and
therefore less signal significance.
Figure~\ref{figFermiDwarfLepton2} shows the \fermilat\ limits for the
DM annihilation cross section for the case of a $\mu^+\mu^-$ final
state \citep{Abdo:2010ex}.
Figure~\ref{figFermiDwarfLepton2}a
assumes photon production only by final-state radiation,
whereas the Figure~\ref{figFermiDwarfLepton2}b
shows the effect, for a single dSph, of including IC scattering.
In the latter interpretation, data from just a single dSph have excluded
much of
the parameter space of DM models devised to explain the \pamela\
and \fermilat\ positron and electron results described 
in Section~\ref{sec:cosmicrays}.

\iact{s} can produce competitive limits for DM annihilation in dSphs only
for very high WIMP masses or with leptonic or $\gamma\gamma$ final states
because of their relatively high energy thresholds.
As illustrated in Figure~\ref{figDMSpectrum}, even for annihilation of
1~TeV WIMPs into $b\bar b$, almost all of the photons are well below 
the \iact{} thresholds.
Of the existing \iact{s}, \magic\ has the lowest threshold
and, therefore, should achieve the best sensitivity to DM.
The \magic\ collaboration has published flux upper limits
above 100~GeV for the dSphs Draco
and Willman~1
from observations made by the first of the two 17-m \magic\
telescopes \citep{Lombardi:2009bh}.
Comparing with predictions from four representative mSUGRA models, the
\magic\ collaboration finds the expected fluxes
to be at least three orders of magnitude below their upper limits.
[Note some caution should be used when considering the limits derived           
for Willman~1, because Willman et al. (arXiv:1007.3499)                  
state that foreground contamination                                             
in their observations and the unusual stellar kinematic distribution mean       
that the DM mass for Willman~1 is not robustly determined.                      
Consequently, the constraints from X-rays and gamma rays that assume            
an equilibrium DM model                                                         
are strongly affected by the systematic uncertainties in their                  
optical observations.]

The \hess\ experiment has published DM limits from observations made on four
dwarf galaxies: Sagittarius, Carina, Sculptor, and Canis Major, of which
the first is classified as a dwarf elliptical and the last as a dwarf
irregular \citep{Aharonian:2007km,Aharonian:2008dm,Glicenstein10}.
The upper limits for cross section, assuming annihilation into
gauge-boson-pair ($W, Z$) final states, have minima
for WIMP masses $M_\chi \sim 1-2$ TeV.
\cite{Glicenstein10} compares the \hess\ Sculptor limits
to the corresponding
limits from the \fermilat\ (a $b\bar b$ final state was assumed in the
latter case, but that does not make
a very large difference in the photon spectrum).
The \hess\ cross section minimum near 2~TeV is nearly equal to an extrapolation
of the \fermilat\ limit, but below that WIMP mass it is not competitive.

\veritas\ has observed four dSphs with energy thresholds ranging
from 300~GeV to 380~GeV: Draco, Ursa Minor, Bootes~1, and
Willman~1 \citep{Acciari:2010pj}, of which only the last was not included
in the \fermilat\ DM limits.
The upper limits show a broad minimum around 1~TeV, with the best cases,
Willman~1 and Ursa Minor,
at $\langle\sigma v\rangle\approx 10^{-23}\,{\rm cm^3 \, s^{-1}}$
(see Figure~\ref{figVeritas_dSph}).
This is a factor of a few higher than the corresponding \fermilat\ upper
limit from Ursa Minor, although an exact comparison is difficult owing to
different assumptions about the composition of the final state
as well as the astrophysical factors.
In general, none of the gamma-ray observations from dSph locations come
close to the most interesting model space
except for the \fermilat\ results for WIMP masses $<100$~GeV.

So far, no DM limits have been derived from neutrino observations of
dwarf galaxies.
The relevance of neutrino observations is thoroughly explored
by \cite{Sandick09} within the context of leptonic models such as
those developed to explain the \pamela\ positron excess.
In short, the \icecube\ experiment will contribute at the very high-mass
end, for example, in the case of WIMPs of mass greater than
7~TeV annihilating to muon pairs.
Of course, it would be the only suitable type of experiment if the WIMPs
were to annihilate or decay primarily to neutrinos!

Detecting DM annihilation or decay in satellite galaxies that are truly
dark or have not yet been detected in the optical is more difficult, both in
execution and interpretation.
Thus far, no DM cross section or lifetime limits have been placed by
collaboration analyses of unidentified sources in \fermilat\ data, although preliminary reports of
a null result from 10 months of observation have been given based on work
in progress \citep{Wang09, Bloom10}.
The difficulty with interpretation is that to predict how many dark
satellites should be detected for a given annihilation cross section and
WIMP mass, for example, requires a very detailed model of the Galactic DM
distribution.
Models based on many-body simulations, such as Via
Lactea-II \citep{Diemand:2008in}, Aquarius \citep{Springel08}, or
GHALO \citep{Zemp09}, are candidates, but multiple realizations of a given
model are needed in order to account statistically for the unknown position
of the Solar System with respect to the DM subhalos.

For this topic, the searching has to be done by an all-sky instrument
such as the \fermilat, but the \iact{s} could play an important role in terms
of follow-up observations of candidates.
Data analyses start from the list of unidentified sources in the
the first \fermilat\ catalog (1FGL), which was derived for the first
11-months of LAT data \citep{Abdo:2010ru}.
In the work presented by \cite{Bloom10}, sources are selected to eliminate
those in the Galactic plane ($|b| \leq 10^\circ$) or those that exhibit
transient behavior.
Then two test statistics are evaluated: one to test whether the source
spectrum is more consistent with DM ($b\bar b$ or $\mu^+\mu^-$ final states)
than with a simple power law, and another to test whether the source is
significantly extended versus being consistent with a point source.
Candidates passing cuts on the test statistics are carefully studied to
avoid backgrounds such as two overlapping point sources.
Background sources can be further reduced by multiwavelength studies of
the candidate locations \citep{Baltz06}.
By this procedure, the analysis reported by \cite{Bloom10} found no dark
satellite candidates in the first 10 months of \fermilat\ data,
but no upper limits were given.
However, estimates of the \fermilat\ sensitivity to dark satellites have
been published elsewhere.
\cite{Baltz08} showed that, for 100~GeV WIMPs annihilating with
the cross section
of Equation~\ref{eqn:standard_cross_section} and a semianalytic model of the
DM distribution, a dozen satellites are expected to be detected
at $5\sigma$ significance or better in a five-year mission.
Another recent study, based on the Via Lactea-II many-body simulation,
is less optimistic, predicting for the same WIMPs between one and five
detections at $3\sigma$ significance in a 10-year mission \citep{Anderson10a}.
However, the same work also predicts that for nearly all of the dark
satellites detected with $> 3\sigma$ significance a non-point-like extension
is expected to be seen at $\gtrsim 5\sigma$ significance, which is
encouraging for analyses that rely heavily on the source extension to
eliminate background.

Analyses that consider only the spectra of unidentified sources can be
interesting, however, not only for small or distant subhalos that cannot
be resolved but also for postulated point-like DM sources, such as DM
concentrations around intermediate mass black holes \citep{Bertone:2005xz}.
\cite{Baltz08} explored the sensitivity of the \fermilat\ to such sources
prior to launch, but that was primarily a consideration of the point-source
sensitivity of the instrument, a topic by now very well studied during
the preparation of the 1FGL \citep{Abdo:2010ru}.
The remaining difficulty is distinguishing such sources from background
via spectral analysis.
Detecting a large population of candidates all fitting a single, large
WIMP mass would be the ideal signal, but that has not happened.
Recently, \cite{Sandick:2010yd} explored this subject without doing
detailed spectral analysis.
They concluded that many (or even all!) of the
368 \fermilat\ unidentified sources at least $10^\circ$ from the Galactic plane
and of at least $5\sigma$ significance could be due to DM.
However, rather than try to put a constraint on DM, they turned the argument
around and attempted to constrain the astrophysics and cosmology.

\subsection{Galaxy Clusters}
\label{sec:galaxyclusters}

\begin{figure}
\begin{center}
\includegraphics[height=2.1in]{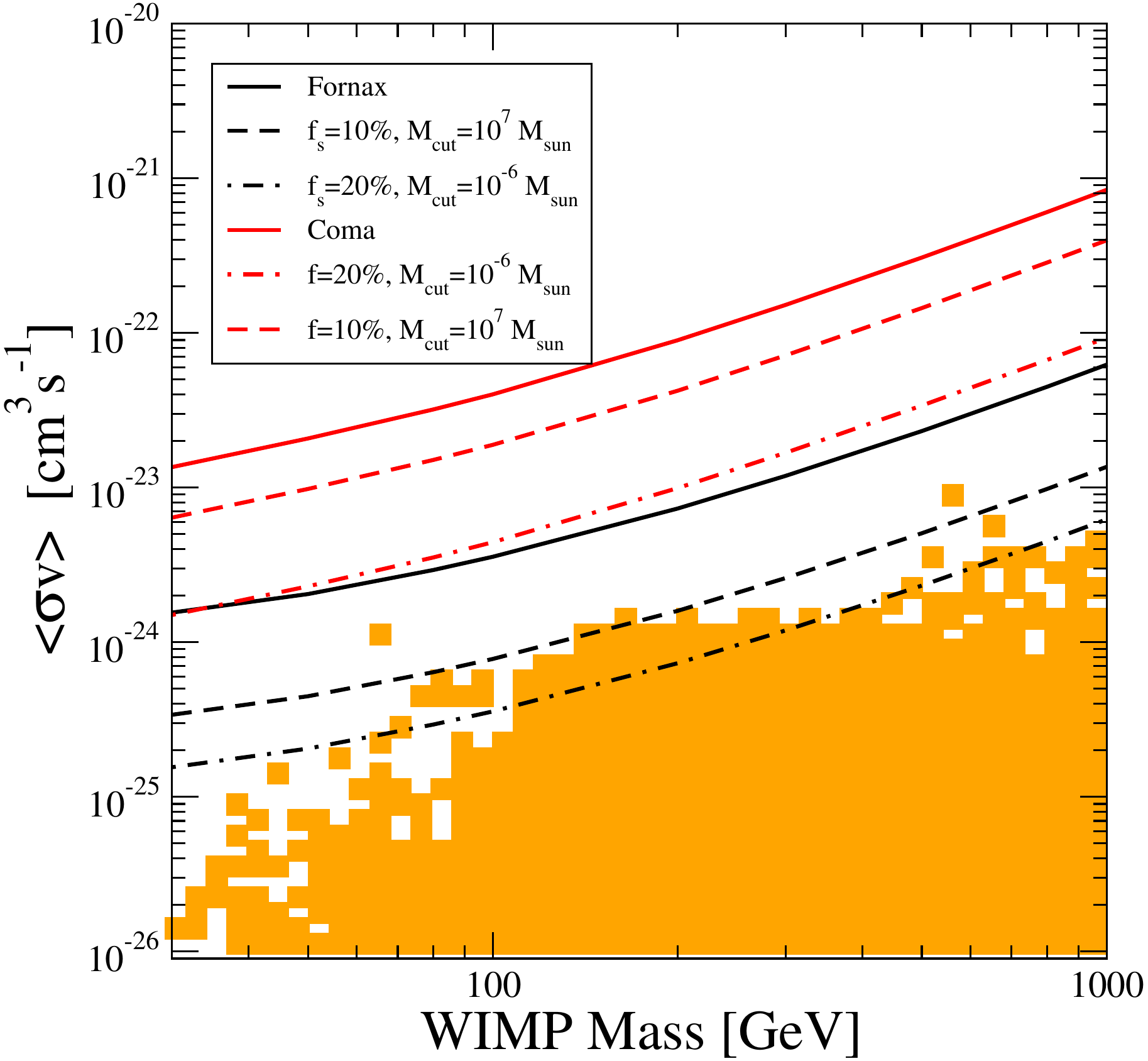}
\includegraphics[height=2.1in]{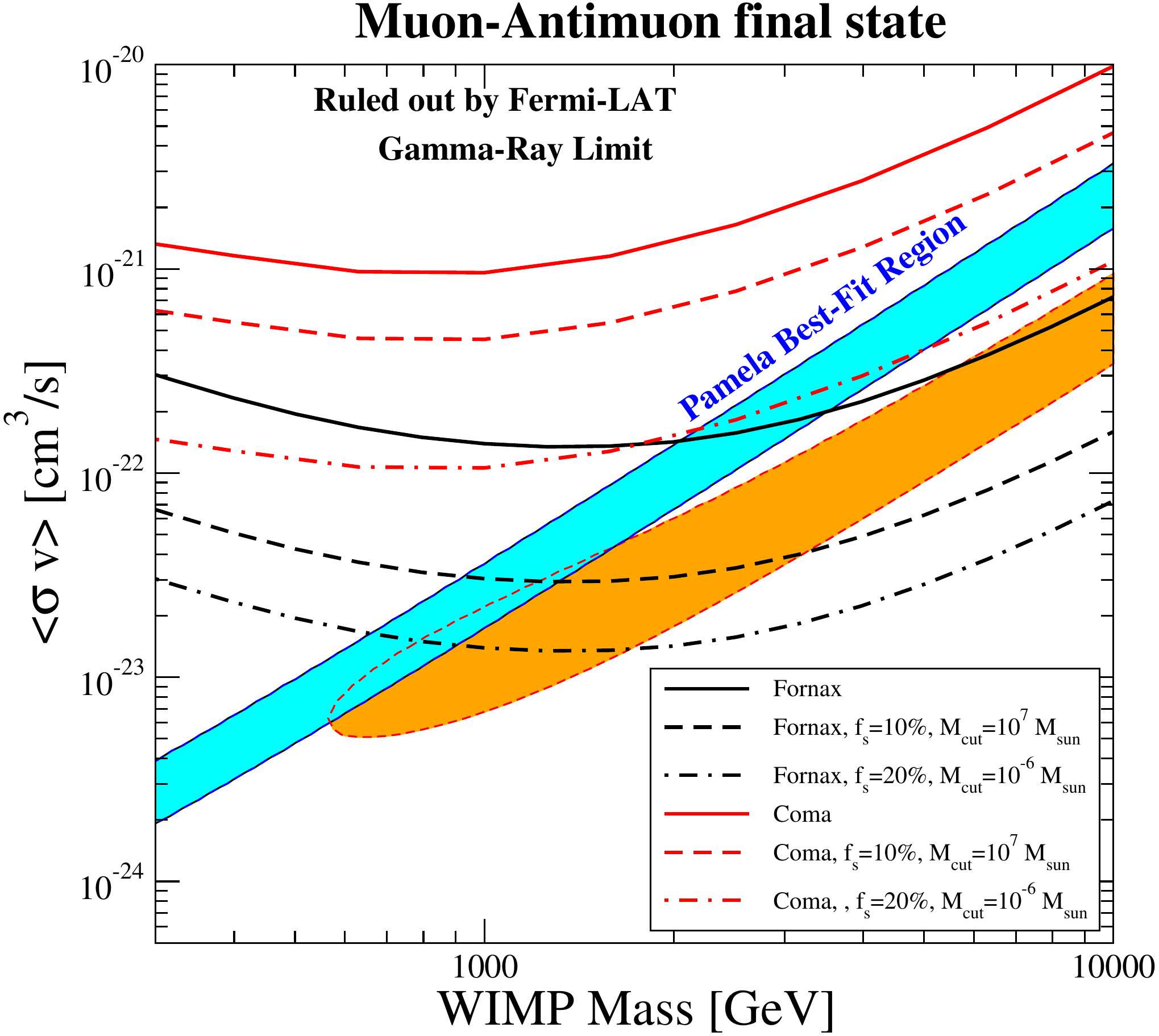}
\caption{\fermilat\ upper limits from \cite{Ackermann:2010rg} for the
dark matter annihilation cross-section for a $b\bar b$ final state (left panel)
and a $\mu^+\mu^-$
final state (right panel) for the Coma and Fornax clusters, including
the effect of substructure on
the expected gamma-ray signal.
The constraints are shown for no
substructure (solid lines), substructure down to the scale of dwarf
galaxies (dashed lines), and substructure down
to $10^{-6} M_\odot$ (dot-dashed lines).
The orange points in the left panel shows the predicted cross section for a 
set of
minimal supersymmetric models that have thermal relic densities compatible
with the observed universal dark matter density, similar to the red 
points in Figure~\ref{figFermiDwarfbbbar}.
The contours
in the right panel show regions allowed by dark matter
models that provide good fits to the
\pamela\ positron fraction (blue) or the \fermilat\ total electron 
spectrum (orange).}
\label{figGalaxyCluster}
\end{center}
\end{figure}

Galaxy clusters are the most massive collapsed structures in the Universe
and are known to be DM dominated.
X-ray studies of the hot intergalactic gas show that not only is there more
mass in the gas than in the galaxies themselves, typically by a factor of
two, but for the gas to be in hydrostatic equilibrium there must be several
times more mass in DM than in the
gas or galaxies \citep[e.g.,][]{Reiprich02,Vikhlinin06}.
Clusters are more distant than local dwarfs, but they are also far more
massive and are therefore attractive targets for DM searches.
However, they also sometimes contain intense gamma-ray sources in the
form of AGN.
The typical angular size of even nearby clusters (a few tenths of a
degree to a degree) is comparable or smaller
than the \fermilat\ PSF except at GeV or
greater energies, so it is difficult or impossible to subtract the
contribution of a single AGN (for \iact{s} the PSF is $\sim 0.1^\circ$,
comparable to that of the thin section of the \fermilat\ $\gtrsim 10$ GeV).
Apart from those containing known AGN or radio galaxies (such as M87 in
the Virgo cluster or NGC 1275 in Perseus), no gamma-ray signal has yet
been associated with a galaxy cluster.
The best flux upper limits below 100~GeV come from the \fermilat, for
33 clusters \citep{Ackermann:2010qj} observed over 11 months, typically at
the level of a few times $10^{-9} \, {\rm cm^2 \, s^{-1}}$.
The \iact{s} CANGAROO, \hess, \magic, and \veritas\ between them have reported
upper limits from observations of six galaxy
clusters \citep{Kiuchi:2009aj,Aharonian:2008uq,Aharonian:2009bc,Aleksic:2009ir,Perkins:2008rw},
three of which
are not included in the \fermilat\ publication.
The flux limits obtained by the \iact{s} for high-mass WIMPs
are three to four orders of magnitude below those of
the \fermilat, reflecting the very large effective areas of the
\iact{s} (albeit with much shorter observation times).
However, the \fermilat\ sensitivity is better for many sources
because of the much higher energy thresholds of the \iact{s}.
For example, in the Perseus cluster the radio galaxy NGC~1275 is
strongly detected by
the \fermilat\ \citep{Abdo:2009wt}, but only upper limits have
followed from \iact{} observations \citep{Aleksic:2009ir,Acciari:2009uq}.

Translating the flux limits to DM limits is more model dependent
than for dSphs observations.
Clusters are expected to include CR proton and electron populations
that would produce gamma rays through collisions
with the intergalactic medium and through IC scattering on the CMB and
intracluser radiation field \citep{Jeltema:2008vu}.
This significantly complicates the interpretation of a signal but may be
ignored when setting conservative DM limits. 
However, the DM spatial distribution is critical to limits for
DM annihilation because the rate is proportional to density squared.
X-ray studies provide a scale radius and DM total mass for each cluster.
Unlike the case of dSphs, for which boosts from substructure are not
expected to be important (and are typically ignored in setting limits, to be
conservative), substructure is likely to be important in clusters from the
galactic scale on down.
At the very least, clusters obviously are full of structure on the galaxy
and dwarf-galaxy scales, so ignoring substructure entirely would
run the risk of setting an overly conservative limit.

The \fermilat\ collaboration
has published DM limits from observations of six galaxy
clusters \citep{Ackermann:2010rg}.
Navarro-Frenk-White profiles are assumed for the DM distributions, with
the scale densities and radii calculated
from the cluster virial masses, as determined from X-ray
observations \citep{Reiprich02} together with the the X-ray
concentration-virial mass relation from \cite{Buote:2007kx}.
Any DM signal from clusters is assumed to be unresolvable by the \fermilat,
as the Navarro-Frenk-White
scale radii lie between only $0.26^\circ$ and $0.45^\circ$, and
the signal would be further concentrated by the density-squared dependence
of the annihilation rate.
Both $b\bar b$ and $\mu^+\mu^-$ final states are considered.
In the latter case, IC scattering of final-state electrons
from the CMB is included and, in fact,
dominates within the \fermilat\ acceptance for WIMP masses above a few
hundred GeV.

With no substructure assumed, the resulting conservative limits are not
as good as those in \cite{Abdo:2010ex} for dSphs.
At 100~GeV they are at best a factor of $\sim 100$
above Equation~\ref{eqn:standard_cross_section}.
Figure~\ref{figGalaxyCluster} shows the annihilation cross section upper
limits from the \fermilat\ for two of the six clusters, the best and worst
cases, including the improvements that result from assuming substructure.
For the best case, which is the Fornax cluster with substructure
down to $10^{-6}M_\odot$, the upper limits are nearly equal to those
obtained from the best-case \fermilat\ dSph observation for $b\bar b$
final states.
The more conservative assumption of substructure down to the dwarf galaxy
level gives upper limits about three times higher in the best case.

Only one of the clusters observed by the \iact{s} (Coma) is included in
the \fermilat\ limits for DM annihilation.
As in the case of dSph observations, the high threshold energies of \iact{s}
prevents them from being sensitive to low-mass WIMPs, especially
for non-leptonic final states.
In fact, the \iact{} publications do not include DM interpretations at all,
except in one case, the \magic\ observation of the Perseus
cluster given by \citet{Aleksic:2009ir}.
A direct comparison to a \fermilat\ result is not possible,
but the conclusion of these researchers
is that a boost (from substructure and/or Sommerfeld
enhancement) of at least $10^4$ is needed in order to be sensitive to a
MSSM model with the highest $f_{\rm SUSY}$ factor, as
defined by \cite{Sanchez-Conde2007}.

\section{OUTLOOK}
\label{sec_summary}

\begin{figure}
\begin{center}
\includegraphics[width=4.5in]{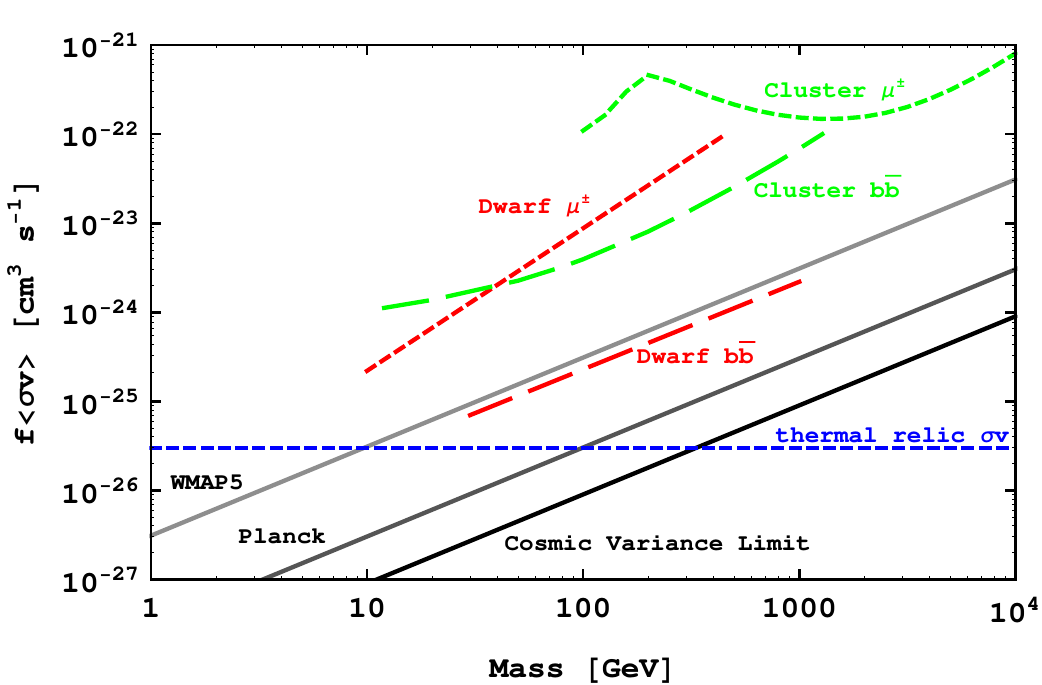}
\caption{
\label{Fig:CMB_constraints}
\small
Constraints on the DM annihilation 
cross section \sigmav from various instruments.
The gamma-ray limits are shown for dwarf spheroid galaxies for $\mu^\pm$
(red short-dashed line) and $b\bar{b}$ (red long-dashed line), and 
galaxy clusters for $\mu^\pm$ (green short-dashed line) and $b\bar{b}$ 
(green long-dashed line) final states, respectively.
The CMB-derived limit (for all annihilation channels except neutrinos) 
using WMAP 5-year data is shown as the light grey line, 
projected limits for Planck are shown as the medium grey line, and the 
dark grey line shows the ultimate cosmic variance limit on the possible 
reach of this technique \citep[reproduced from][]{Galli:2009zc}.
For the CMB limits, $f$ parameterizes the coupling of the 
annihilation products to the gas
in the early Universe and is usually $\mathcal{O}(1)$.
The dashed blue line shows the thermal self-annihilation cross section given
by Equation~\ref{eqn:standard_cross_section}.}
\end{center}
\end{figure}

As of the date of this review, there has been no definitive detection
of a DM signal in astroparticle data. 
This does not mean that no CRs or gamma rays from DM have been detected.
Rather, given
the astrophysical uncertainties associated with interpreting the data, 
where a signal has been detected, 
there is insufficient evidence to allow unambiguous attribution 
to DM annihilation or decay scenarios.
Thus far, the astroparticle data
have been most useful for limiting the properties of proposed DM 
particles in the contexts of specific models.
In this review, we discussed the most recent experimental results 
together with the related astrophysical phenomenology to provide an 
understanding of issues associated with interpreting the data. 
Understanding these uncertainties and how they can be quantified 
and/or reduced is essential
to further progress in this area of research.

The PAMELA rising positron abundance remains intriguing, but its 
interpretation is ambiguous.  
DM explanations for these data, whether invoking WIMP decay or 
annihilation, have not been completely ruled out by recent gamma-ray data.
However, given the non-standard nature of these DM models and 
the existence of
reasonable alternative astrophysical explanations, the DM explanation
is inconclusive.

AMS-02 \citep{Bindi:2010zz} is expected to launch in early 2011.  
It will significantly improve on current CR observations,
especially for positrons and antiprotons.  
For example, the enhanced particle-identification capabilities of AMS-02 
will clarify
whether the PAMELA positron measurements are contaminated by
misidentified protons, and they will extend the positron data
to higher energies.  
This is likely to improve the indirect search
for DM in general \citep{Beischer:2009zz} and the interpretation
of the PAMELA positron abundance in particular.  

Future results from gamma-ray telescopes, the \fermilat\ in particular, 
will show
significantly improved sensitivity to DM annihilation or decay.
Searches for DM in the halo of the Milky Way with the \fermilat\ data are
underway.
DM halo searches have the advantage of large numbers of signal 
photons, but their 
potential will only
be realized, in terms of approaching sensitivity to the most interesting 
models, if
better constraints can be put on the large Galactic diffuse background.
That will require better understanding of the propagation of CRs 
in the Galaxy as well as
improved knowledge of the gas distributions and ISRF.
It has long been expected that a DM signal may be detectable 
toward the GC because the annihilation rate should peak
in this region.
But, 
the numerous point sources, together with the inter-relationship
with the diffuse emission and its uncertainties along the line-of-sight, 
and instrumental effects make analysis complicated.
As for the halo searches, improvements in understanding the foreground 
emission, together with
source identification, are crucial for this kind of analysis.
However, 
the DM limits from observations of dwarf spheroidal satellite 
galaxies are sure to improve steadily as the \fermilat\ accumulates 
up to 10~times 
more data than have been used to set limits so far.
The statistical power will further improve as
more sophisticated analyses make simultaneous fits to the ensemble of dwarfs 
and as more dwarfs are discovered by upcoming optical surveys that will 
greatly improve the sky coverage
(e.g. DES \citep{Abbott:2005bi}, BigBOSS \citep{Schlegel:2009uw}, 
Pan-STARRS, LSST \citep{Abel:2009pq}).
Further stellar-velocity observations of dwarfs are also likely to improve 
knowledge of the astrophysical factors (Equation~\ref{eqn:dwarfFlux}).
Therefore, 
it is reasonable to expect that by the end of the \fermilat\ mission the 
dwarf analyses will be sensitive
to annihilation of WIMPs into heavy quarks or gauge bosons
at or below the standard annihilation cross
section, Equation~\ref{eqn:standard_cross_section},
for WIMP masses up to at least $100\,{\rm GeV}\,c^{-2}$.

Of course, within the time frame of the \fermilat\ mission, the 
Large Hadron Collider experiments
will explore the parameter space in certain particle physics models 
for production of WIMPs up to much
higher masses than $100\,{\rm GeV}\,c^{-2}$ \citep{Hooper:2008sn}.
If such a particle is discovered, then the indirect detection experiments will
not only be important for studying its possible role as a DM particle,
but they may also have much more information to work with, in particular 
the WIMP mass and strengths of couplings to the standard model.

In the near term, the CMB also provides a method 
for detecting or constraining particle DM that is potentially competitive 
with the astroparticle searches.
Because the number of DM particles
remains constant after freeze-out, their number density
varies as $n \propto z^3$, where $z$ is the redshift.
Because the annihilation rate per 
unit volume is $\sim n^2 \left<\sigma v\right>$,
annihilation is greatly enhanced at early times.
In particular, the annihilation rate around the time of
recombination, $z \sim 1000$, could have been large enough to affect the 
CMB significantly.
If the products of DM annihilation
were anything besides neutrinos, they would have
coupled to the surrounding gas and caused significant reionization.
This leads to a relatively strong constraint from current WMAP data,
which exclude DM with the standard annihilation cross
section (Equation~\ref{eqn:standard_cross_section}) and
mass $\lesssim 10\,{\rm GeV}\,c^{-2}$.
The current limit using this method is shown in 
Figure~\ref{Fig:CMB_constraints}, which also summarizes the limits derived
from gamma-ray data discussed earlier.
Future data from the Planck satellite \citep{Tauber10} will cover a 
larger region of WIMP parameter
space, either detecting or excluding DM with the
standard annihilation cross section up to $\sim 100\,{\rm GeV}\,c^{-2}$.
For a more thorough discussion see 
\citet[][and references therein]{Galli:2009zc}.
Along with gamma-ray observations, 
measurements of the CMB by instruments like Planck 
constitute one of the most interesting possibilities
for detecting or constraining the properties of DM in the near future.

\section*{Acknowledgements}

We are grateful for the assistance and comments by the following people:
B. Anderson, W. Atwood, S. Digel, T. Jeltema, P. Michelson, 
I. Moskalenko, S. Murgia, 
E. Murphy, J. Nielsen, S. Profumo, S. Ritz, and L. Strigari.
We acknowledge the assistance of E. Mocchiutti and P. Picozza in the
preparation of Figure~\ref{figCRPAMELA}.



\end{document}